\documentclass{article}
\usepackage{arxiv}

\usepackage[utf8]{inputenc}
\usepackage[T1]{fontenc}
\usepackage[numbers,sort&compress]{natbib}
\usepackage{doi}
\usepackage[dvipsnames,svgnames]{xcolor}
\definecolor{paperlinkblue}{rgb}{0.01,0.31,0.65}
\usepackage{hyperref}
\newcommand{\figref}[1]{\hyperref[#1]{Figure~\ref*{#1}}}
\newcommand{\panelref}[2]{\hyperref[#1]{Figure~\ref*{#1}#2}}
\newcommand{\figtworef}[2]{\hyperref[#1]{Figures~\ref*{#1}} and~\hyperref[#2]{\ref*{#2}}}
\newcommand{\tabref}[1]{\hyperref[#1]{Table~\ref*{#1}}}
\newcommand{\secref}[1]{\hyperref[#1]{Section~\ref*{#1}}}
\newcommand{\suppfigref}[1]{\hyperref[#1]{Supplementary Figure~\ref*{#1}}}
\newcommand{\supppanelref}[2]{\hyperref[#1]{Supplementary Figure~\ref*{#1}#2}}
\newcommand{\supptabref}[1]{\hyperref[#1]{Supplementary Table~\ref*{#1}}}
\usepackage{url}
\usepackage{booktabs}
\usepackage{amsfonts}
\usepackage{nicefrac}
\usepackage{float}
\usepackage{microtype}
\usepackage{xspace}
\usepackage{graphicx}
\usepackage{placeins}
\usepackage{amsmath}
\usepackage{amssymb}
\usepackage{array}
\usepackage{colortbl}
\usepackage{longtable}

\newcommand{\ouralgo}{\textsc{PRISMA-LLM}\xspace}
\newcolumntype{L}[1]{>{\raggedright\arraybackslash}p{#1}}
\newcolumntype{C}[1]{>{\centering\arraybackslash}p{#1}}
\definecolor{PrismaHeader}{HTML}{66649A}
\definecolor{PrismaBand}{HTML}{FFFCCB}
\definecolor{PrismaLLMPurple}{HTML}{2D176F}

\graphicspath{{figures/main_manuscript/}{figures/publication_analysis/}{figures/supplementary/}}

\title{\ouralgo: An Empirical Reporting Framework for AI-Assisted Systematic Reviews}

\author{Miguel Zabaleta$^1$, Baihan Lin$^{1,2,3,*}$  \\ \\
  $^1$ Department of AI and Human Health, Icahn School of Medicine at Mount Sinai, New York, NY, USA \\
  $^2$ Department of Psychiatry, Icahn School of Medicine at Mount Sinai, New York, NY, USA \\
  $^3$ Department of Neuroscience, Icahn School of Medicine at Mount Sinai, New York, NY, USA \\
  $^*$ Corresponding: \texttt{baihan.lin@mssm.edu}
}

\renewcommand{\headeright}{}
\renewcommand{\undertitle}{Manuscript}
\renewcommand{\shorttitle}{\ouralgo: An Empirical Reporting Framework for AI-Assisted Systematic Reviews}

\hypersetup{
pdftitle={\ouralgo: An Empirical Reporting Framework for AI-Assisted Systematic Reviews},
pdfauthor={Miguel Zabaleta and Baihan Lin},
pdfkeywords={systematic reviews, large language models, PRISMA, reporting guidelines, evidence synthesis, review automation},
colorlinks=true,
linkcolor=paperlinkblue,
citecolor=paperlinkblue,
urlcolor=paperlinkblue
}

\begin{document}
\maketitle
\begingroup
\renewcommand{\thefootnote}{}
\begin{NoHyper}
\footnotetext{Analysis code, publication-date data used in the analyses, PRISMA-LLM reporting instructions, a fillable checklist workbook, and the checklist evidence table are available at \url{https://github.com/linlab/PRISMA-LLM}.}
\end{NoHyper}
\endgroup

\begin{abstract}
Large language models (LLMs) and AI-enabled software increasingly participate in systematic-review decisions, yet the information needed to audit these workflows is reported inconsistently. We analyze SciLitBench, a corpus of 888 review-automation papers with 14,726 annotations, to characterize changes in methods, review-stage use, evaluation and reported limitations. Automation has shifted toward LLM- and software-facing workflows, including stages that can alter the evidence base. Since 2023, 38.0\% of software/product papers reported no evaluation, compared with 9.3\% of LLM papers. Reporting coverage increased with LLM workflow complexity, yet 52\% of positive-only LLM evaluations still reported an unmet reliability or performance requirement. From these patterns, we introduce PRISMA-LLM, an empirically grounded framework separating implementation disclosure from consequence-sensitive evaluation and limitation reporting.
\end{abstract}



Systematic reviews help determine which evidence enters scientific syntheses and, in many fields, inform guidelines, policy, and clinical decisions. Computational systems have long assisted literature search and screening, but LLMs and AI-enabled software are now being used across a broader portion of the review workflow, including full-text screening, data extraction, quality assessment, and synthesis \citep{tsafnat2014systematicReviewAutomation,marshall2019towardAutomation,scherbakov2025emergence}. As automation moves from navigation toward decisions that can alter the evidence base, the reproducibility of a review increasingly depends on details of the computational workflow itself.

This creates a reporting problem. An AI-assisted workflow can depend on model or product identity, version and access mode, prompts and system instructions, sampling settings, schemas, document parsing, chunking, retrieval, post-processing, thresholds, and human adjudication. For hosted or proprietary systems, behavior may also change over time without a corresponding change to the published review. Without sufficient disclosure, readers may be unable to determine what was automated, how an error could propagate, whether a decision was independently checked, or whether the workflow can be rerun or meaningfully audited \citep{nationalacademies2019reproducibility}.

Reporting guidelines turn general expectations of transparency into concrete information that readers can use. PRISMA 2020 specifies the information needed to understand how a systematic review was conducted and reported \citep{page2021prisma}; PRISMA-S provides a close precedent by recognizing that technically consequential search procedures require dedicated reporting detail \citep{rethlefsen2021prismaS}. AI-assisted review workflows create a related reporting problem: without the model and workflow details, readers lack the information needed to know what was automated, how errors could arise, or whether the workflow can be reused. Several recent proposals have begun to address AI-related reporting, including PRISMA-AI, PRISMA-trAIce, PRISMA-DFLLM, and L-PRISMA \citep{cacciamani2023prismaAi,holst2025prismaTraice,susnjak2023prismaDfllm,shailendra2026lPrisma}. What remains unclear is how reporting practice has evolved across the broader review-automation literature and which recurring dimensions of evaluation and limitation reporting can provide an empirical basis for more specific guidance.

Here we analyze SciLitBench \citep{zabaleta2026scilitbench}, a corpus of 888 papers on computational automation in systematic reviews. We ask three questions: how review automation has changed over time and across review stages; whether evaluation and limitation reporting differs across methods; and what these observed reporting patterns imply for transparent reporting of LLM-assisted reviews. We then introduce \ouralgo as an empirically informed reporting framework aligned with the structure of PRISMA 2020. Its purpose is not to certify the quality or safety of an AI workflow, but to make the implementation choices, evaluation evidence, human oversight, and failure modes sufficiently visible for readers to interpret and audit the review. Because implementation complexity is not equivalent to methodological risk, we treat the proposed levels as disclosure tiers rather than risk tiers and explicitly allow consequential uses of otherwise simple tools to trigger stronger evaluation expectations.

\section{Results}

SciLitBench provides a longitudinal view of how computational systems are entering systematic-review practice. We first characterize changes in publication volume, technical approach, review-stage use, and model family, then examine whether evaluation and limitation reporting has changed with the methods being used. Finally, we use those empirical patterns to motivate a reporting framework. 

\subsection{Growth and shifts}

\begin{figure}[tbp]
\centering
\begin{minipage}{0.96\textwidth}
\textbf{a) Corpus composition}\par
\vspace{4pt}
\centering
{\small
\setlength{\tabcolsep}{8pt}
\renewcommand{\arraystretch}{1.12}
\begin{tabular}{@{}lr@{\hspace{2.3em}}lr@{}}
\toprule
\textbf{Corpus measure} & \textbf{Count} & \textbf{High-level domain} & \textbf{Papers, n (\%)} \\
\midrule
Paper-level records & 888 & Life sciences and medicine & 548 (61.7) \\
Annotation items & 14,726 & Engineering and technology & 140 (15.8) \\
Evaluation items & 6,562 & Social sciences and management & 115 (13.0) \\
Limitation items & 2,896 & Mixed & 62 (7.0) \\
& & Natural sciences & 21 (2.4) \\
& & Arts and humanities & 2 (0.2) \\
\bottomrule
\end{tabular}}
\end{minipage}

\vspace{12pt}

\begin{minipage}{0.96\textwidth}
\textbf{b) Publication growth}\par
\vspace{2pt}
\centering
\includegraphics[width=\textwidth]{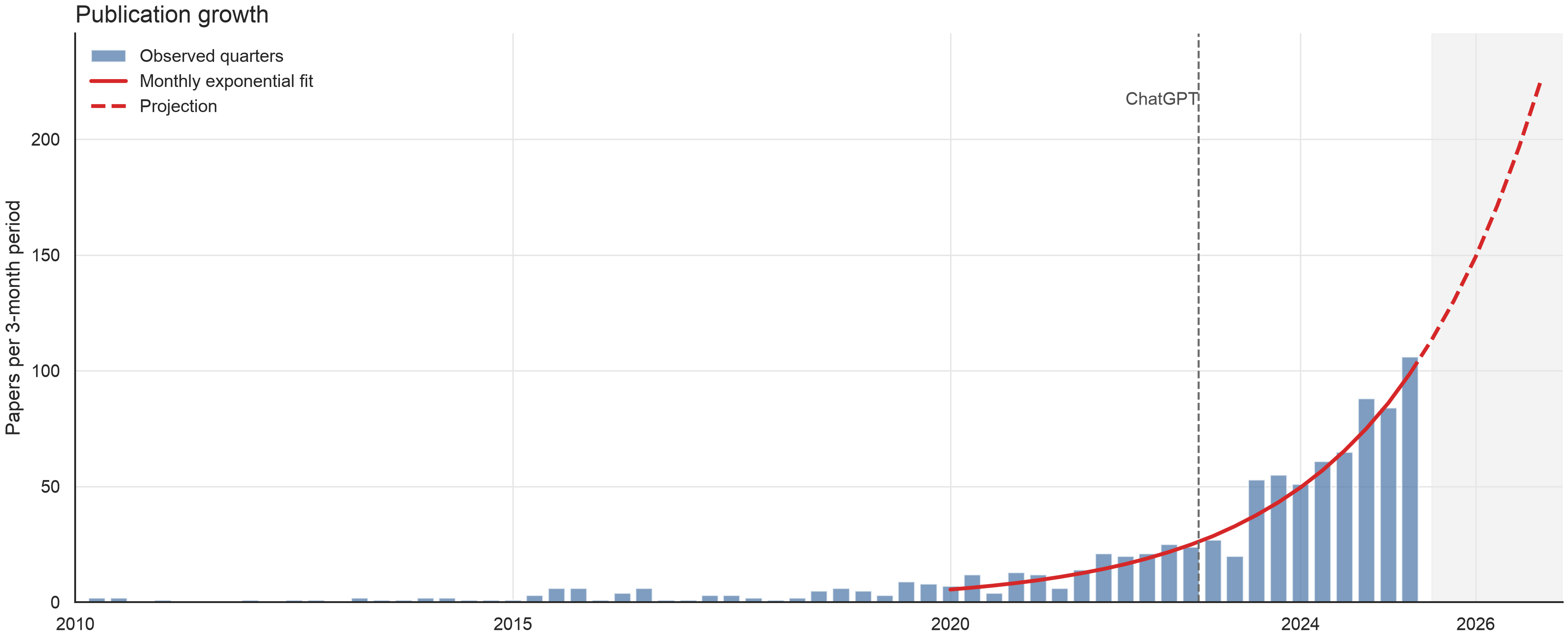}
\end{minipage}
\caption{\textbf{Corpus scope and publication growth in review automation.} \textbf{a)} SciLitBench contains 888 paper-level records and 14,726 annotation items; 61.7\% of papers are from life sciences and medicine. \textbf{b)} Publication counts are shown from 2010 onward for legibility; five included papers predate 2010. The red curve is a log-linear fit to positive monthly counts from January 2020 through June 2025, aggregated to the displayed periods. The dashed continuation and grey region show a descriptive extrapolation through 2026, not a causal or calibrated forecast. The vertical dashed line marks the public release of ChatGPT.}
\label{fig:growth}
\label{fig:corpus-growth}
\end{figure}

Publication volume increased sharply over the observed period. \panelref{fig:growth}{b} shows that publication counts were low and relatively flat for most of the corpus history, began to rise around 2018–2019, and then steepened after the release of ChatGPT at the end of 2022. A log-linear fit to positive monthly counts from January 2020 through June 2025 corresponds to an estimated 4.7\% monthly growth rate. The corpus spans several fields, with 61.7\% of records coming from life sciences and medicine (\panelref{fig:growth}{a}). This concentration fits the broader history of evidence synthesis where biomedical research has generated a large systematic-review literature (PRISMA itself grew out of reporting guidance for reviews and meta-analyses of health-care interventions) \citep{page2016epidemiologyBiomedicalReviews,hoffmann2021nearly80,moher2009prismaStatement,page2021prisma}. Engineering and technology and social sciences and management add 15.8\% and 13.0\%. Mixed-domain papers, natural sciences, and arts and humanities make up the rest.

\begin{figure}[tbp]
\centering

\begin{minipage}{0.90\textwidth}
\textbf{a) Approach orientation}\par
\vspace{2pt}
\centering
\includegraphics[width=\linewidth]{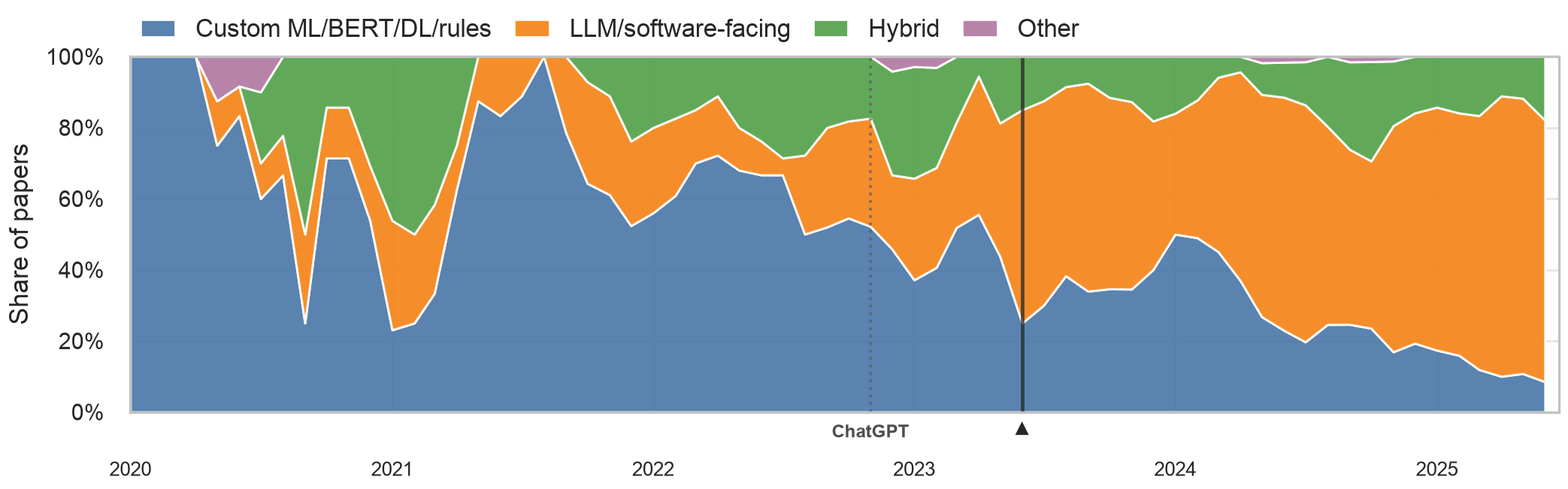}
\end{minipage}

\vspace{5pt}

\begin{minipage}{0.90\textwidth}
\textbf{b) Review-stage composition}\par
\vspace{2pt}
\centering
\includegraphics[width=\linewidth]{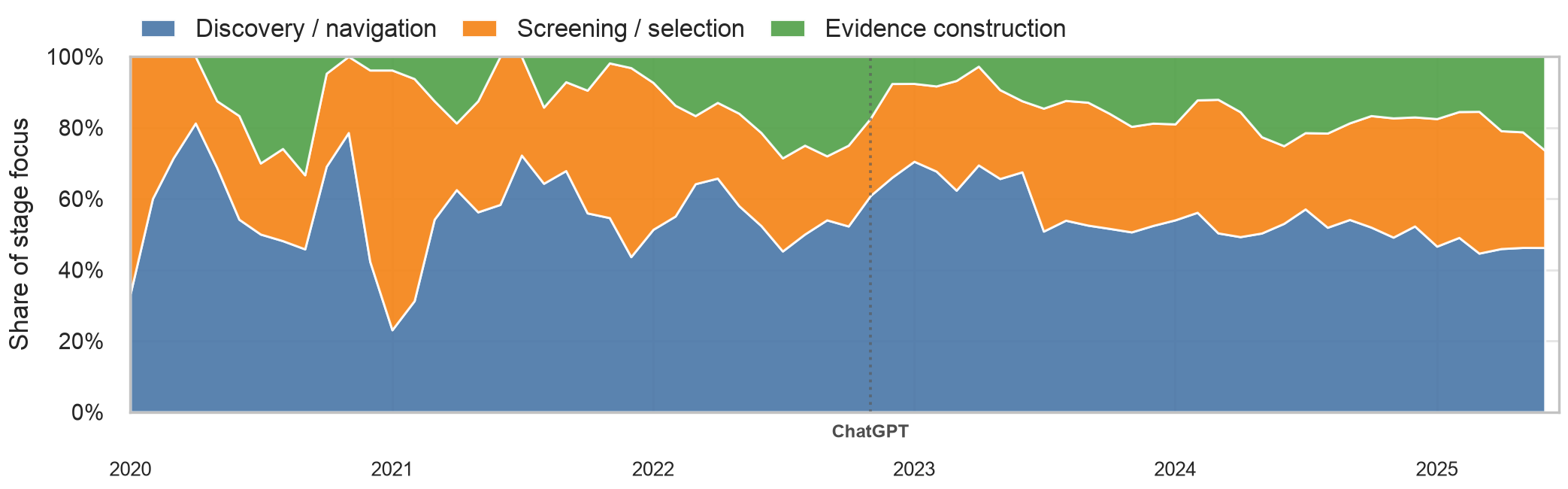}
\end{minipage}

\vspace{5pt}

\begin{minipage}{0.90\textwidth}
\textbf{c) LLM adoption by domain}\par
\vspace{2pt}
\centering
\includegraphics[width=\linewidth]{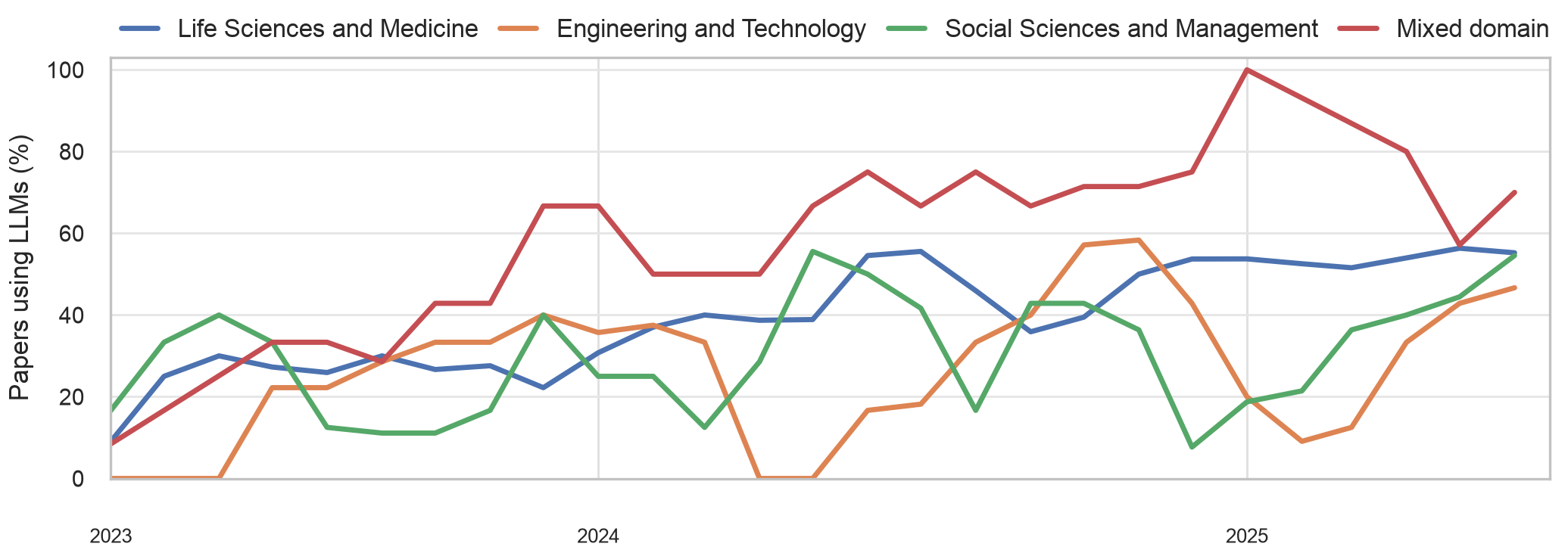}
\end{minipage}

\caption{\textbf{Shifts in method orientation, review-stage use, and LLM adoption.} All panels use overlapping rolling 3-month windows. \textbf{a)} Exclusive approach orientation. The LLM/software-facing category includes LLM and software-product papers; before ChatGPT, this category is therefore software-facing. The dotted line marks ChatGPT's public release, and the solid line and triangle mark the first window in which LLM/software-facing work became the largest orientation category. \textbf{b)} Review-stage composition. Each paper contributes total weight one, divided equally across its assigned stage groups; screening/selection and evidence construction therefore represent the share of review-stage focus, not the share of unique papers. \textbf{c)} Share of papers using an LLM within each high-level domain, beginning with the first window containing LLM use. Mixed-domain papers span more than one domain and are shown separately. Adjacent points share observations and should not be interpreted as independent time points.}
\label{fig:method-stage-shifts}
\label{fig:llm-adoption-domain}
\end{figure}

\figref{fig:method-stage-shifts} shows how review stages and methods have shifted over time. Earlier review-automation work was centered on custom model-building, BERT, deep learning, traditional machine learning, and rule-based systems. In recent years, we see a clear transition to LLMs and software products that use LLMs under the hood, like Elicit and DistillerSR \citep{byun2023elicitLanguageModels,distillerSrPlatform} (\panelref{fig:method-stage-shifts}{a}). Automation is also moving into review stages where errors can change the evidence base. We group these stages as discovery/navigation, screening/selection, and evidence construction. In recent records, screening/selection and evidence construction account for a substantial share alongside discovery/navigation (\panelref{fig:method-stage-shifts}{b}). 

We also examine how LLM adoption differs across fields. In life sciences and medicine, the share of papers using LLMs has risen, approaching 60\% in recent months (\panelref{fig:method-stage-shifts}{c}). Engineering and technology and social sciences and management show more short-term variation, but both curves rise through the observed 2025 window after a late-2024 to early-2025 dip. Mixed-domain records reach 70--80\% in several recent windows.

We also explored whether disciplines differ in the language models they use. \suppfigref{fig:supp-model-family} shows that BERT-based papers have more domain specialization: biomedical variants dominate much of life sciences and medicine, while generic BERT, BERTopic, sentence embeddings, and other transformer variants appear more prominently outside that field. Most LLM papers use GPT/ChatGPT-family systems, consistent with prior reports \citep{scherbakov2025emergence,lieberum2025largeLanguageModelsSystematicReviews}, with Claude, Gemini/Bard/PaLM, Llama, Mistral/Mixtral, and other families appearing less often. As for the use of open-weight versus proprietary models, the data shows that most LLM usage still relies on proprietary or hosted systems, while open-weight models appear much less often (84.1\% proprietary, 11.0\% mixed, and 4.9\% open-weight). Even though we see more open-weight usage in some domains than others (for example, mixed-domain records include open-weight or mixed proprietary/open-weight use in 28.0\% of cases, compared with 10.3\% in engineering and technology), these differences are not statistically significant.

Overall, the corpus shows a field changing on several fronts: review-automation papers are growing rapidly, the methods used have shifted toward LLM/software-facing workflows, and automation is increasingly present in stages that can affect which evidence enters a review and how it is interpreted. Next, we explore whether reporting practice has kept pace with these changes, and introduce reporting richness as a measure of how much evaluation and limitation detail each paper provides.

\subsection{Evaluation reporting differs by automation context and method complexity}

Review automation is growing and moving into parts of the review where automated decisions can affect the evidence base. We therefore asked whether papers give researchers enough detail to understand what was automated, how it was evaluated, and what limits were reported. We summarize this with reporting richness, a descriptive index of paper-level coverage rather than a measure of study quality, and compare it across approach groups and LLM-era workflow types.

Reporting richness is a descriptive 0--15 index of how much evaluation and limitation detail a paper reports. It combines five dimensions: performance evidence, comparisons, optimization or modification evidence, resource or feasibility information, and limitations or failure modes; the full scoring rule is given in the Methods (\secref{sec:derived-measures}). Traditional, deep-learning, and BERT-oriented approaches decline as a share of recent papers while their mean reporting richness remains comparatively stable. LLM papers show increasing reporting richness after 2023, whereas software/product papers remain common but contain less paper-level evaluation and limitation detail (\suppfigref{fig:supp-evaluation-reporting}). Since 2023, software/product papers average a reporting-richness score of 3.3, compared with 6.3 for LLM papers and 6.0 for traditional, deep-learning, and BERT-oriented papers. The share of papers with no reported evaluation shows the same contrast: 38.0\% of software/product papers report no evaluation, compared with 9.3\% of LLM papers and 13.6\% of traditional, deep-learning, and BERT-oriented papers.

\begin{figure}[!t]
\centering

\begin{minipage}{0.96\textwidth}
\textbf{a) Method-complexity mix}\par
\vspace{2pt}
\centering
\includegraphics[width=\linewidth]{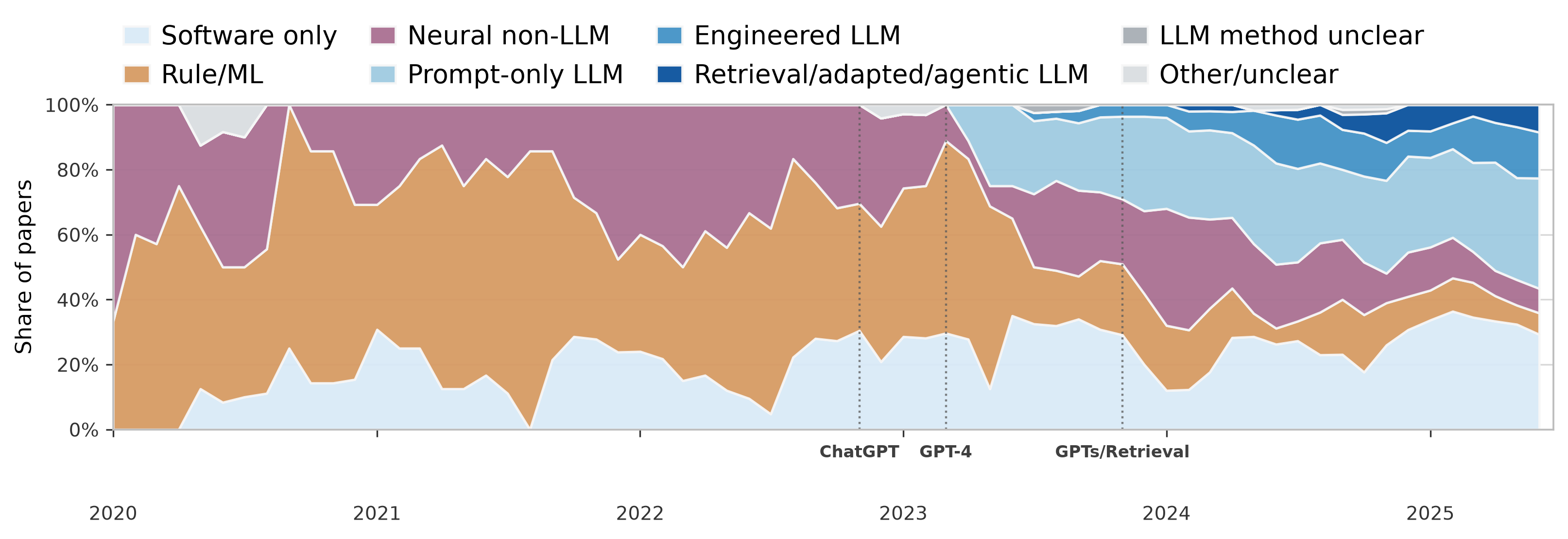}
\end{minipage}

\vspace{6pt}

\begin{minipage}{0.96\textwidth}
\textbf{b) Reporting by method complexity}\par
\vspace{2pt}
\centering
\includegraphics[width=\linewidth]{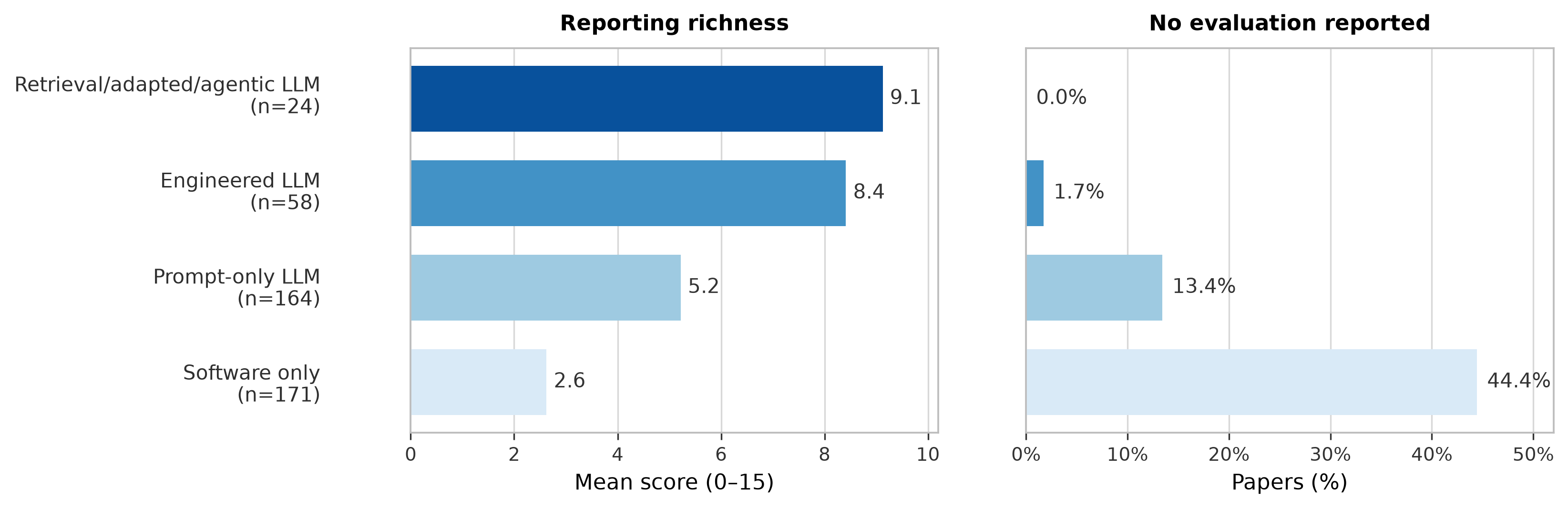}
\end{minipage}

\caption{\textbf{LLM-era workflow complexity and evaluation reporting.} \textbf{a)} Exclusive method-complexity categories in overlapping rolling 3-month windows. Reference dates for ChatGPT, GPT-4, and GPTs/Retrieval provide temporal context only and are not treated as causal breakpoints. \textbf{b)} Reporting outcomes among papers published from 2023 onward. Bars show mean reporting richness and the percentage of papers with no reported evaluation; $n$ is the number of papers in each group. Reporting coverage increases across the observed LLM workflow categories, from prompt-only to engineered and retrieval/adapted/agentic methods. These are descriptive associations and may also reflect differences in study purpose, venue, task, and other characteristics.}
\label{fig:reporting-complexity}
\label{fig:complexity-mix}
\label{fig:complexity-reporting}
\end{figure}

The same corpus also lets us look inside the methods used in LLM-era review automation. Rather than treating all LLM use as one category, we grouped papers by the kind of automation method reported: software/product-only use, prompt-only LLM use, structured or engineered LLM workflows, retrieval, adapted, or agentic LLM workflows, and non-LLM automation. \panelref{fig:complexity-mix}{a} shows the rolling 3-month share of these groups alongside three timing references: ChatGPT, GPT-4, and GPTs/Retrieval. Prompt-only LLM use first appears in 2023, structured or engineered workflows become visible later that year, and retrieval, adapted, or agentic workflows appear in early 2024. In the latest rolling window, ending June 2025, the split is 34.0\% prompt-only LLM, 29.2\% software/product only, 14.2\% engineered or structured LLM, 8.5\% retrieval, adapted, or agentic LLM, 7.5\% neural non-LLM, and 6.6\% rule or traditional ML.

Among papers published from 2023 onward, software/product-only papers have the lowest mean reporting-richness score (2.6), and 44.4\% report no evaluation. Prompt-only LLM papers average 5.2, with 13.4\% reporting no evaluation; engineered or structured LLM papers average 8.4, with 1.7\% reporting no evaluation; and retrieval, adapted, or agentic LLM papers average 9.1, with no papers in this group annotated as reporting no evaluation (\figref{fig:complexity-reporting}). The monotonic pattern is descriptive rather than causal: more complex workflows are often themselves the object of method development, whereas off-the-shelf software may be used instrumentally in papers whose primary contribution lies elsewhere.

Together, these results motivate \ouralgo, a reporting framework designed to make implementation choices, human oversight, evaluation evidence, and failure modes more visible.

\subsection{Usefulness and adequacy are distinct judgments}
 
\begin{figure}[tbp]
\centering
 
\begin{minipage}{0.96\textwidth}
\raggedright
\textbf{a) Overall impression among LLM papers over time}\par\smallskip
 
\includegraphics[width=\linewidth]{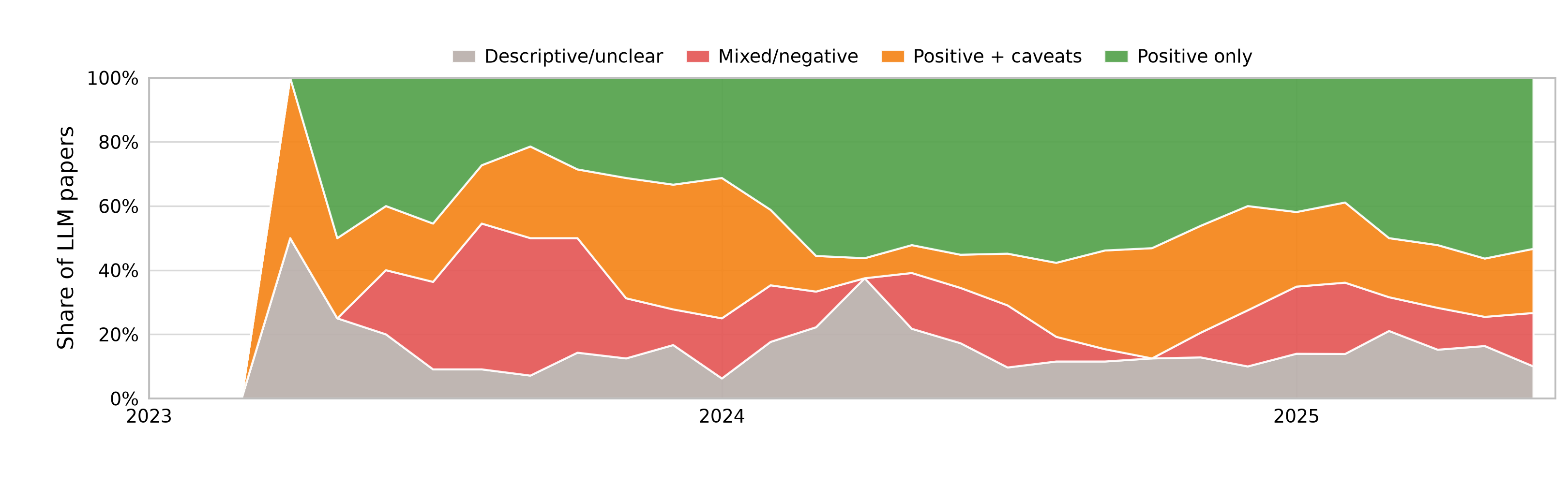}
\end{minipage}
\vspace{0.8em}
 
\begin{minipage}{0.96\textwidth}
\raggedright
\textbf{b) High-bar concerns by overall impression}\par\smallskip
 
\includegraphics[width=\linewidth]{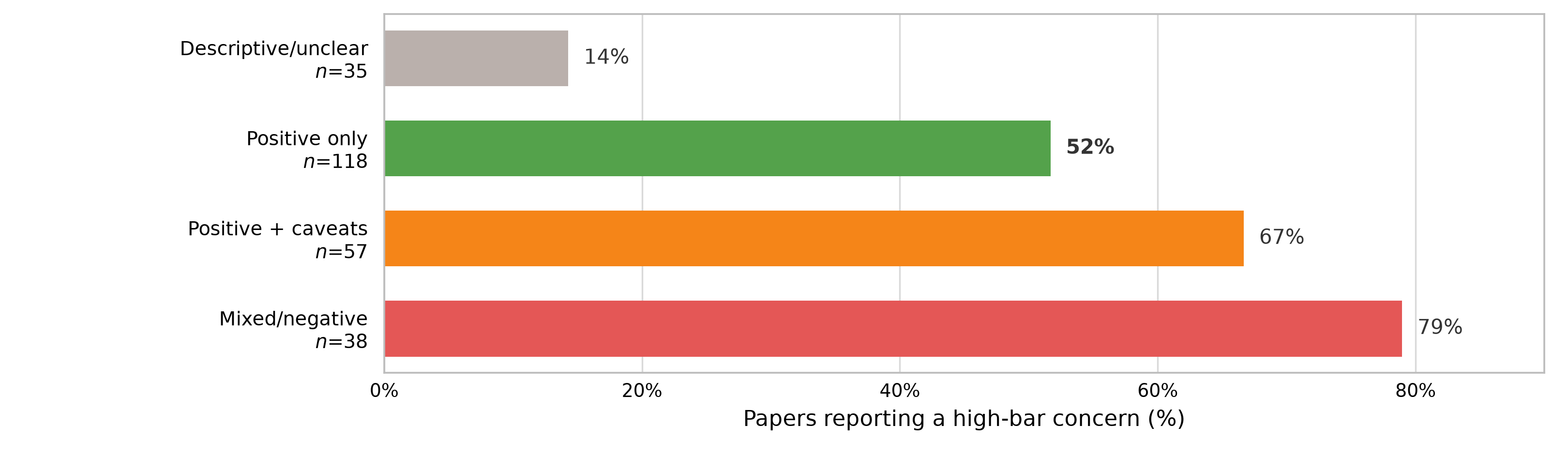}
\end{minipage}
 
\caption{\textbf{Positive assessments frequently coexist with unmet reliability requirements.} \textbf{a)} Shares of LLM papers assigned to four mutually exclusive overall-impression categories within rolling 3-month windows. Adjacent windows contain overlapping papers, and the earliest windows contain few records. \textbf{b)} Percentage of papers in each impression category that report at least one concern that the workflow remains below the reliability or performance bar required for its intended use; $n$ denotes the number of papers in each category. High-bar concerns appear across all impression categories. Notably, 52\% of papers classified as positive only still report at least one such concern.}
 \label{fig:utility-adequacy}
 \end{figure}

Overall evaluations of LLM-assisted review automation were often favorable, but favorable assessments did not imply that authors considered the systems adequate for unsupervised use. Among 118 LLM papers classified as positive only, 52\% nevertheless reported at least one limitation indicating that the workflow remained below the reliability or performance bar required for its intended role. The corresponding share was 67\% among 57 positive-with-caveats papers and 79\% among 38 mixed/negative papers (\figref{fig:utility-adequacy}). These categories capture reported judgments rather than independent re-evaluation of the systems, but they reveal a recurring distinction between usefulness and adequacy for delegation. A workflow may save time or perform well on average while still requiring human verification at points where errors could change the evidence base.

\subsection{\ouralgo}

\ouralgo is an evidence-informed reporting framework for LLM-assisted and AI-enabled systematic-review workflows, organized to complement PRISMA 2020. Its empirical basis is the recurring evaluation and limitation dimensions observed in SciLitBench and the descriptive association between workflow type and reporting coverage \citep{zabaleta2026scilitbench}.  Its implementation requirements additionally draw on the workflow choices encountered in SciLitBench and on PRISMA 2020, PRISMA-S, and adjacent AI-reporting proposals.

PRISMA 2020 provides the baseline reporting structure for the review itself, while PRISMA-S demonstrates how a technically consequential component of the review can require dedicated reporting detail \citep{page2021prisma,rethlefsen2021prismaS}. PRISMA-AI focuses on reviews and meta-analyses of AI systems in healthcare; PRISMA-trAIce addresses AI-tool identification, human--AI interaction, performance evaluation, and limitations; PRISMA-DFLLM emphasizes domain-specific LLMs, fine-tuning, reproducibility, and ethical considerations; and L-PRISMA adds reporting for GenAI-assisted screening and summarization \citep{cacciamani2023prismaAi,holst2025prismaTraice,susnjak2023prismaDfllm,shailendra2026lPrisma}. \ouralgo differs from these proposals by deriving its graded reporting levels from observed patterns in a large corpus of review-automation papers. Its reporting scale is motivated by the relationship we observe between method type, reporting richness, and workflow complexity in SciLitBench. \ouralgo is intended to complement existing efforts by linking a detailed implementation-disclosure layer to reporting dimensions in an evidence-based approach. 

The framework has three layers (\figref{fig:prisma-llm-architecture}). First, a universal checklist specifies the information that should be reported across LLM-assisted review workflows (\tabref{tab:prisma-llm-main-checklist}). Second, five implementation-disclosure levels organize the additional technical detail needed to reconstruct workflows from off-the-shelf tools to coordinated multi-agent systems. Third, consequence-sensitive evaluation expectations determine when validation should exceed the minimum associated with a disclosure level. The levels are therefore \emph{disclosure tiers, not risk tiers}. A technically simple workflow can still be highly consequential if it autonomously excludes records or supplies unverified extracted evidence; conversely, a technically elaborate workflow used only for non-substantive drafting may have limited influence on the evidence base. Task consequence, degree of human verification, system opacity and reversibility should therefore be reported explicitly and can justify stronger evaluation at any implementation level. The cumulative minimum reporting dimensions associated with the five implementation levels are summarized in \suppfigref{fig:prisma-llm-reporting-map}.

\begin{figure}[H]
\centering
\fbox{\begin{minipage}{0.94\textwidth}
\small
\textbf{Layer 1: universal reporting checklist.} Every LLM-assisted review should identify the system and task, describe inputs and outputs, report human oversight, disclose evaluation and failures, and provide reproducibility materials where possible (\tabref{tab:prisma-llm-main-checklist}).

\medskip
\textbf{Layer 2: implementation-disclosure level.}
\begin{center}
\begin{tabular}{@{}C{0.09\textwidth}L{0.22\textwidth}L{0.58\textwidth}@{}}
\toprule
Level & Workflow & Additional implementation detail \\
\midrule
1 & Off-the-shelf software & Product/version, interface, role, settings, validation status and availability constraints. \\
2 & Prompt-only LLM & Level 1 plus model identity, prompts/system instructions, inference settings, repetitions and output constraints. \\
3 & Structured/engineered & Level 2 plus examples, schemas/rubrics, prompt chains, voting/ensembles, LLM-as-judge and post-processing. \\
4 & Retrieval/adapted/single-agent & Level 3 plus retrieval sources, parsing/chunking, embeddings, adaptation/fine-tuning, thresholds, agent actions and failure handling. \\
5 & Multi-agent/swarm & Level 4 plus agent roles, coordination, handoffs, termination, shared state, conflict resolution and cross-agent audit logs. \\
\bottomrule
\end{tabular}
\end{center}

\medskip
\textbf{Layer 3: consequence-sensitive evaluation.} Minimum evaluation breadth increases with implementation complexity, but implementation level is not a risk score. At any level, stronger validation is expected when AI outputs can change search retrieval, inclusion/exclusion, extracted evidence, risk-of-bias judgments, synthesis or conclusions; when human verification is limited; when the system is proprietary or mutable; or when errors are difficult to reconstruct or reverse.
\end{minipage}}
\caption{\textbf{The three-layer structure of \ouralgo.} The universal checklist specifies what every report should disclose; the implementation-disclosure level specifies the technical details needed to reconstruct the workflow; and task consequence, human verification, opacity and reversibility determine whether evaluation should exceed the minimum associated with the implementation level. This separates implementation complexity from methodological risk.}
\label{fig:prisma-llm-architecture}
\end{figure}

\begin{table}[!tbp]
\centering
\scriptsize
\setlength{\tabcolsep}{3pt}
\renewcommand{\arraystretch}{1.10}
\begin{tabular}{@{}L{0.12\textwidth}L{0.15\textwidth}L{0.66\textwidth}@{}}
\toprule
Section & Item & What to report \\
\midrule
Identification & LLM-1: Title/abstract & Identify substantive LLM or AI-enabled software use and the review stage(s) affected; distinguish review decisions/evidence processing from writing-only assistance. \\
Rationale & LLM-2: Purpose & Explain why AI was used and which review burden or methodological problem it was intended to address. \\
Methods & LLM-3: System identification & Report system/tool name, model/version, provider, access date, interface/API, deployment mode, and whether access was proprietary, open-weight, local, hosted or custom. \\
Methods & LLM-4: Task and disclosure level & Specify each review task, assign the PRISMA-LLM implementation-disclosure level, and state whether outputs could alter retrieval, inclusion/exclusion, extraction, appraisal, synthesis or conclusions. \\
Methods & LLM-5: Inputs/document processing & Describe records, abstracts, PDFs, tables, supplements, examples, labels and prior decisions supplied to the system, including parsing, OCR, section/table handling, chunking, context limits, truncation and retrieval. \\
Methods & LLM-6: Prompts/settings & Report prompts and system instructions, examples, schemas/rubrics, inference settings, repetitions/sampling, output constraints and model-specific settings needed to reproduce the workflow. \\
Methods & LLM-7: Outputs/post-processing & Describe generated labels, rankings, extracted fields, judgments or text; output schemas; parsing/format repair; thresholds; confidence scores; and how model outputs became review decisions or analysis data. \\
Methods & LLM-8: Workflow development & Report consequential prompt, schema, retrieval or orchestration changes, the rationale for the final workflow, and important failed or alternative configurations that informed it. \\
Methods & LLM-9: Task-specific workflow & For screening, extraction, appraisal, annotation or LLM-as-judge use, report task-specific decision rules, reference standards, calibration, thresholds, edge-case handling and adjudication procedures. \\
Methods & LLM-10: Human oversight & State who reviewed AI outputs, what proportion was independently checked, how disagreements/corrections were handled, and who retained final responsibility for consequential decisions. \\
Methods & LLM-11: Evaluation plan & Pre-specify performance, comparative, modification/optimization, resource/feasibility, and limitation/failure-mode evaluation as applicable; strengthen evaluation when task consequence, limited verification, opacity or irreversibility warrants it. \\
Results & LLM-12: AI/human processing counts & Report numbers of records, reports, fields, annotations, claims or judgments processed by AI, humans or both, including how automation altered the flow of records through the review. \\
Results & LLM-13: Evaluation and errors & Report task-specific performance and required comparative/modification/resource evidence, plus false positives/negatives, hallucinations, extraction/parsing failures, disagreements, corrections and audit outcomes. \\
Discussion & LLM-14: Limitations and implications & Explain model/version dependence, prompt sensitivity, validation limits, proprietary opacity, remaining human workload and how observed or plausible failures could affect the evidence base or conclusions. \\
Availability & LLM-15: Reproducibility materials & Provide prompts, system instructions, settings, schemas, code, validation samples, labels, audit logs and raw/parsed outputs when legally and ethically possible; report data-sharing, copyright, privacy, vendor and terms-of-service constraints. \\
Disclosure & LLM-16: Support and interests & Report sponsored access, credits, private model access, vendor involvement, material support and relevant competing interests. \\
\bottomrule
\end{tabular}
\par\vspace{6pt}
\caption{\textbf{PRISMA-LLM reporting checklist for LLM-assisted and AI-enabled systematic reviews.} This page-sized checklist is the primary user-facing framework and is intended to be used alongside PRISMA 2020 rather than as a replacement for it. The expanded item wording is provided in \supptabref{tab:prisma-llm-checklist}. Implementation-level instructions, a fillable checklist workbook, and the checklist evidence table are available at \url{https://github.com/linlab/PRISMA-LLM}.}
\label{tab:prisma-llm-main-checklist}
\end{table}


\section{Discussion}

Across 888 papers, review automation is becoming more common and moving into stages where computational outputs can determine which studies are retained and what evidence is extracted from them. At the same time, the information available to evaluate those workflows is uneven. Software/product papers were especially likely to omit paper-level evaluation, whereas LLM papers reported progressively more evaluation detail as workflows became more engineered. These findings identify a practical accountability gap: adoption can be straightforward even when the evidence needed to audit the adopted system is sparse.

A second finding is that favorable assessments and unmet reliability requirements frequently coexist. More than half of LLM papers with positive-only overall evaluations nevertheless reported at least one concern that the workflow remained below a performance or reliability bar for its intended use. This distinction matters because ``useful'' is not the same claim as ``adequate for delegation.'' A tool may reduce workload, improve recall, or support extraction while still requiring human verification at exactly the points where errors could alter the review's evidence base. Reporting should therefore make both utility and residual failure modes visible, together with who retained final responsibility for consequential decisions.

The observed association between workflow complexity and reporting richness provides an empirical rationale for graded disclosure, but complexity should not be mistaken for risk. The present categories capture implementation choices needed to reconstruct a workflow; they do not measure the consequence of an error, the amount of human oversight, or the opacity of a proprietary system. An off-the-shelf product used to autonomously exclude records may require more stringent validation than a multi-agent system used only to draft text that is fully checked by authors. We therefore position the proposed \ouralgo levels as disclosure tiers and add an explicit consequence override: workflows that can change the evidence base should report task-specific validation and human verification regardless of level.

The study also has limitations. SciLitBench is concentrated in life sciences and medicine, covers publications through June 2025, and reflects the search and inclusion criteria of the parent systematic review. Reporting richness counts annotated evidence items and captures breadth with limited within-dimension depth; it is not a validated study-quality score. The observed complexity gradient is descriptive and may be confounded by study purpose, venue, task, publication year, or the fact that engineered workflows are more often themselves the object of method-development studies. Paper-level absence of evaluation also does not prove that no validation exists elsewhere, for example in a product report, protocol, repository, or prior publication. Finally, the proposed checklist combines empirical observations with workflow experience and prior reporting guidance; it has not yet undergone formal consensus development or prospective user testing.

We introduce PRISMA-LLM as an empirically grounded reporting framework proposed in this study. It is not presently an official extension endorsed by the PRISMA Executive. Formal consensus development, independent usability testing and prospective evaluation can test and refine the framework as AI-assisted review practice evolves.

When AI systems participate in evidence synthesis, the computational path from source documents to review decisions should be inspectable. This requires reporting what the system received and produced, what humans checked, where failures occurred, and how those failures could have affected the evidence base. \ouralgo translates these requirements into a checklist, implementation-disclosure scheme and consequence-sensitive evaluation framework for testing and revision.

\section{Methods}

\subsection{Study design and corpus}

This study is a secondary analysis of the 888 full-text papers included in the SciLitBench corpus \citep{zabaleta2026scilitbench}. SciLitBench was built from a systematic review of computational methods used to automate literature reviews. Its librarian-informed search covered PubMed, Semantic Scholar, and Scopus and yielded 42,981 candidate records after database export and deduplication. The corpus includes papers published through June 2025. The upstream screening workflow identified 1,819 title-and-abstract inclusions, retrieved 1,012 full texts, and produced 888 final inclusions after full-text screening and duplicate removal. Each included paper serves as one unit of analysis.

The corpus contains six annotated fields: publication year, domain, review stage, computational approach, evaluation results, and limitations. Each paper has one high-level and one medium-level domain label. Review stage and computational approach are multi-label fields. The evaluation-results and limitations fields contain item-level evidence statements, including quotations preserved from the source paper. The harmonized analysis inventory contains 14,726 annotation items across the 888 papers: 888 year values, 1,776 domain labels, 1,350 review-stage labels, 1,254 approach labels, 6,562 evaluation-result items, and 2,896 limitation items (\supptabref{tab:supp-corpus-inventory}).

\subsection{Publication dates and temporal analysis}

The source annotations provided publication year. We recovered publication month for all 888 papers using bibliographic databases, publisher records, and dates reported in source documents. The final date table contains one unique paper identifier and a recovered month and year. A separate override file documents 33 manually adjudicated dates and the evidence or decision basis for each. All temporal analyses end in June 2025, so the 2025 results represent January through June rather than a complete publication year.

Monthly plots use rolling 3-month windows. For each displayed month, paper counts were summed over that month and the preceding two months before calculating proportions. Adjacent points therefore contain overlapping papers. Domain-level LLM adoption was calculated as the number of papers annotated with an LLM approach divided by the total number of papers in that domain and rolling window. For review-stage composition, each paper received a total weight of one, divided equally across its review-stage groups, before the rolling proportions were calculated. Approach-orientation and method-complexity plots use one exclusive group per paper, as defined below.

Publication growth was calculated from unique papers per publication month. The exponential curve in \panelref{fig:growth}{b} was fitted to positive monthly counts from January 2020 through June 2025 by ordinary least squares on the log count,
\[
\log(N_m)=\alpha+\beta m.
\]
The fitted monthly values were summed into 3-month calendar periods for display. The monthly growth rate was calculated as $\exp(\beta)-1$, and the 2026 extrapolation was calculated as the sum of the 12 fitted monthly values. This extrapolation was used only to summarize the recent trajectory; it was not treated as a calibrated forecast or evidence of a causal break at the release of any model or product.

\subsection{Review-stage, method, and model groupings}

Review-stage labels were grouped according to their role in the review. The discovery and navigation group includes search or retrieval, topic modeling or clustering, and text mining or information synthesis. The screening and selection group includes title-and-abstract screening, full-text screening, and screening used for ranking or prioritization. The evidence construction group includes data extraction, risk-of-bias or quality assessment, and claim verification. A paper may contribute to more than one review-stage group.

We used both multi-label approach groups and an exclusive approach orientation. The multi-label groups are traditional/deep-learning/BERT, LLM, and software/product. Papers containing more than one approach can contribute to more than one of these groups. The exclusive orientation classifies a paper as custom ML/BERT/deep-learning/rule-oriented, LLM/software-facing, hybrid, or other/unclear. Hybrid papers contain at least one custom approach and at least one LLM or software/product approach. We used these definitions for approach trends and reporting comparisons since 2023.

High-level domain labels were used for domain comparisons. Under the final domain schema, papers spanning multiple substantially different domains or describing their corpus too broadly for one ASJC assignment were annotated as \texttt{unknown / not reported}. We display this group as \texttt{Mixed} in the analysis.

Model-family values were normalized from the approach annotations. BERT-family labels were grouped as biomedical variants, generic BERT, BERTopic, scientific embedding variants, sentence embedding variants, general transformer variants, or other BERT-like models. LLM-family labels were grouped as GPT/ChatGPT, Claude, Gemini/Bard/PaLM, Llama, Mistral/Mixtral, other open-weight or open-family models, other proprietary or hosted models, or unclear. The model-family panels include papers from 2015 onward; the arts and humanities category was omitted because it contained only two papers, and unclear LLM families were omitted from the LLM-family panel.

For the access analysis, models such as GPT, Claude, and Gemini were classified as proprietary or hosted, while models such as Llama, Mistral, and Qwen were classified as open-weight or open-family. Papers containing both groups were labeled mixed. Model access was coded only when the family could be identified.

\subsection{Reporting measures and empirical method complexity}
\label{sec:derived-measures}

The five dimensions used for reporting richness were derived from SciLitBench's evaluation-result and limitation annotations. The SciLitBench annotation team constructed starting schemas from 50 sampled papers and applied them to the full 888-paper corpus with Llama 3.3-70B. The team then reviewed source quotations and adjudicated the final categories and mappings. This process consolidated recurring evidence into four evaluation dimensions: performance, comparisons, modifications or optimization, and resources or feasibility, plus one dimension for limitations or failure modes. Full details of the annotation schemas, prompts, harmonization procedures, and annotation-quality checks are provided in the SciLitBench paper \citep{zabaleta2026scilitbench}.

Reporting richness is a descriptive paper-level index with a maximum possible score of 15. It sums five dimension scores: performance evidence, comparative evidence, optimization or modification evidence, resource or feasibility evidence, and limitation or failure-mode reporting. Each dimension contributes 0 points for no items, 1 point for one item, 2 points for two or three items, and 3 points for four or more items. The score captures reporting breadth and a limited measure of within-dimension depth; study quality requires separate appraisal.

Performance evidence is the number of items categorized as performance versus human annotation; comparative evidence is the number of comparative-result items; modification evidence is the number of feature- or modification-effect items; and resources or feasibility is the number of runtime-and-cost items. The limitation dimension counts all limitation items except \texttt{none reported}. Overall qualitative impressions, \texttt{none reported}, and \texttt{other} evaluation items do not contribute to the score. Papers were classified as having no evaluation when their evaluation annotations contained the \texttt{none reported} category.

For the supplementary sentiment analysis, overall-evaluation mappings were combined at the paper level. Papers with positive mappings and no mixed or negative mapping were labeled positive only. Papers with both positive and mixed or negative mappings were labeled positive with caveats. Papers containing mixed or negative mappings without a positive mapping were labeled mixed/negative; the remainder were descriptive or unclear. A paper was marked as reporting an unmet high bar when at least one limitation item was categorized as \texttt{high bar needed but not achieved}.

The empirical method-complexity groups in \figref{fig:complexity-mix} were assigned from the approach categories and the normalized method labels attached to LLM annotations. The highest applicable LLM group took precedence. Retrieval, adapted, or agentic workflows required an explicit retrieval method such as RAG, model adaptation such as fine-tuning, or agentic orchestration. Engineered or structured workflows included techniques such as few-shot learning, structured prompting, voting or ensembles, and LLM-as-evaluator designs. Prompt-only workflows used generic or zero-shot prompting without one of these additional methods. Plain zero-shot use was therefore prompt-only, while zero-shot chain-of-thought was engineered. The prompt-only group was restricted to papers with an identified generic or zero-shot method; unresolved cases were labeled LLM method unclear. Non-LLM papers were labeled neural non-LLM when they contained BERT or deep learning, rule/traditional ML when they contained traditional machine learning or rule-based approaches, and software/product only when software was the only substantive approach category. Remaining papers were labeled other/unclear. These groups are analytical categories for the corpus and are separate from the five-level PRISMA-LLM proposal. The executable mapping is available with the analysis code.

\subsection{Development of PRISMA-LLM}

We developed PRISMA-LLM from three inputs: the reporting patterns observed in the SciLitBench corpus, the implementation choices encountered while building SciLitBench, and a comparison with PRISMA 2020 and adjacent reporting proposals for search methods and AI-assisted reviews. The empirical layer uses the five reporting dimensions above and the observed tendency for more complex LLM workflows to report more evaluation and limitation detail. The implementation layer covers choices that can change the behavior of an automated review workflow, including model or tool identity, prompts and settings, schemas, document processing, retrieval and chunking, thresholds, post-processing, human oversight, adjudication, and reproducibility materials. We mapped each checklist item to the empirical findings, workflow experience, or prior reporting guidance that motivated it (\supptabref{tab:supp-evidence}).

The proposed implementation scale contains five levels. Level 1 covers off-the-shelf tools or software; Level 2 covers prompt-only LLM use; Level 3 covers few-shot, structured, engineered, voting, ensemble, or LLM-as-judge workflows; Level 4 covers retrieval-augmented, adapted, fine-tuned, or single-agent workflows; and Level 5 covers multi-agent or swarm systems. These levels summarize implementation complexity rather than methodological risk. The evaluation expectations provide a minimum floor that accumulates across levels, while consequential tasks can trigger stronger requirements regardless of level. Reporting richness remains a descriptive empirical index because no compliance cutoff has undergone prospective validation.

The implementation checklist was mapped to PRISMA 2020 and informed by related reporting guidance, including PRISMA-S, PRISMA-AI, PRISMA-trAIce, PRISMA-DFLLM, and L-PRISMA. The framework is fully specified here; formal Delphi consensus development and prospective user testing are additional validation steps rather than prerequisites for scientific evaluation of the proposal.

\subsection{Statistical analysis}

The paper is the unit of analysis. Counts and percentages are descriptive unless a statistical test is named. Multi-label approach and review-stage percentages use papers as denominators and may sum to more than 100\%. For reporting-richness comparisons, we report arithmetic means and the proportion of papers with no evaluation annotation. Where specified, comparisons by approach and empirical method complexity included only papers published from 2023 onward.

Differences in proprietary versus open-weight LLM use across high-level domains were evaluated after excluding mixed-access and unannotated papers. A global chi-square statistic was compared with 5,000 label permutations using a fixed random seed of 20260630. Pairwise domain comparisons used two-sided Fisher's exact tests, with Benjamini-Hochberg correction across domain pairs. Statistical significance was assessed at a two-sided $\alpha=0.05$.

For exploratory analyses of reported performance evidence, metric names and nearby numeric values were parsed from performance-versus-human quotations. Sensitivity was normalized to recall. Percentages and values greater than 1 and no greater than 100 were converted to the 0--1 scale; fractions and values outside that range were excluded. Multiple values for the same paper, model class, stage group, and metric were collapsed to their median. We excluded LLM/BERT hybrid papers from direct model-class comparisons because paper-level annotations often leave the source model for individual values unresolved. Reported values were summarized by the median and interquartile range. We calculated two-sided Mann-Whitney $U$ tests with tie correction only when each comparison group contained at least five paper-level values. These metric summaries are exploratory inventories of heterogeneous reported evidence rather than controlled benchmark comparisons, and no multiplicity correction was applied to them.

\section*{Code and materials availability}
 
Analysis code, the PRISMA-LLM reporting instructions, a fillable checklist workbook, and the checklist evidence table are available at \url{https://github.com/linlab/PRISMA-LLM}. Access to the underlying SciLitBench corpus and source-derived materials follows the terms described in the SciLitBench resource.

\section*{Acknowledgments}

BL is supported by the U.S. National Institutes of Health (NIH), the U.S. Department of Veterans Affairs (VA), the Brain \& Behavior Research Foundation (BBRF), the Sidney R. Baer, Jr. Foundation, the Hasso Plattner Foundation, and the Windreich Family Foundation. No funding was specific to this work.

\bibliographystyle{sn-nature}
\bibliography{main}

@article{tsafnat2014systematicReviewAutomation,
  author = {Tsafnat, Guy and Glasziou, Paul and Choong, Miew Keen and Dunn, Adam and Galgani, Filippo and Coiera, Enrico},
  title = {Systematic Review Automation Technologies},
  journal = {Systematic Reviews},
  volume = {3},
  pages = {74},
  year = {2014},
  doi = {10.1186/2046-4053-3-74},
  url = {https://doi.org/10.1186/2046-4053-3-74}
}

@article{marshall2019towardAutomation,
  author = {Marshall, Iain J. and Wallace, Byron C.},
  title = {Toward Systematic Review Automation: A Practical Guide to Using Machine Learning Tools in Research Synthesis},
  journal = {Systematic Reviews},
  volume = {8},
  pages = {163},
  year = {2019},
  doi = {10.1186/s13643-019-1074-9},
  url = {https://doi.org/10.1186/s13643-019-1074-9}
}

@misc{byun2023elicitLanguageModels,
  author = {Byun, J. and Stuhlm{\"u}ller, Andreas},
  title = {Elicit: Language Models as Research Tools},
  howpublished = {In OECD, \emph{Artificial Intelligence in Science: Challenges, Opportunities and the Future of Research}},
  year = {2023},
  doi = {10.1787/a8d820bd-en},
  url = {https://www.oecd.org/en/publications/artificial-intelligence-in-science_a8d820bd-en/full-report/elicit-language-models-as-research-tools_fec8a6ab.html}
}

@misc{distillerSrPlatform,
  author = {{DistillerSR}},
  title = {The DistillerSR Platform},
  howpublished = {\url{https://www.distillersr.com/products/distillersr-systematic-review-software}},
  note = {Accessed 2026-07-06}
}

@article{scherbakov2025emergence,
  author = {Scherbakov, Dmitry and Hubig, Nina and Jansari, Vinita and Bakumenko, Alexander and Lenert, Leslie A.},
  title = {The Emergence of Large Language Models as Tools in Literature Reviews: A Large Language Model-Assisted Systematic Review},
  journal = {Journal of the American Medical Informatics Association},
  volume = {32},
  number = {6},
  pages = {1071--1086},
  year = {2025},
  doi = {10.1093/jamia/ocaf063},
  url = {https://doi.org/10.1093/jamia/ocaf063}
}

@article{lieberum2025largeLanguageModelsSystematicReviews,
  author = {Lieberum, Judith-Lisa and Toews, Markus and Metzendorf, Maria-Inti and Heilmeyer, Felix and Siemens, Waldemar and Haverkamp, Christian and Boehringer, Daniel and Meerpohl, Joerg J. and Eisele-Metzger, Angelika},
  title = {Large Language Models for Conducting Systematic Reviews: On the Rise, but Not Yet Ready for Use---A Scoping Review},
  journal = {Journal of Clinical Epidemiology},
  volume = {181},
  pages = {111746},
  year = {2025},
  doi = {10.1016/j.jclinepi.2025.111746},
  url = {https://doi.org/10.1016/j.jclinepi.2025.111746}
}

@book{nationalacademies2019reproducibility,
  author = {{National Academies of Sciences, Engineering, and Medicine}},
  title = {Reproducibility and Replicability in Science},
  publisher = {National Academies Press},
  address = {Washington, DC},
  year = {2019},
  doi = {10.17226/25303},
  url = {https://doi.org/10.17226/25303}
}

@article{moher2009prismaStatement,
  author = {Moher, David and Liberati, Alessandro and Tetzlaff, Jennifer and Altman, Douglas G. and {The PRISMA Group}},
  title = {Preferred Reporting Items for Systematic Reviews and Meta-Analyses: The PRISMA Statement},
  journal = {PLoS Medicine},
  volume = {6},
  number = {7},
  pages = {e1000097},
  year = {2009},
  doi = {10.1371/journal.pmed.1000097},
  url = {https://doi.org/10.1371/journal.pmed.1000097}
}

@article{page2016epidemiologyBiomedicalReviews,
  author = {Page, Matthew J. and Shamseer, Larissa and Altman, Douglas G. and Tetzlaff, Jennifer and Sampson, Margaret and Tricco, Andrea C. and Catal{\'a}-L{\'o}pez, Ferr{\'a}n and Li, Lun and Reid, Emma K. and Sarkis-Onofre, Rafael and Moher, David},
  title = {Epidemiology and Reporting Characteristics of Systematic Reviews of Biomedical Research: A Cross-Sectional Study},
  journal = {PLoS Medicine},
  volume = {13},
  number = {5},
  pages = {e1002028},
  year = {2016},
  doi = {10.1371/journal.pmed.1002028},
  url = {https://doi.org/10.1371/journal.pmed.1002028}
}

@article{hoffmann2021nearly80,
  author = {Hoffmann, Falk and Allers, Katharina and Rombey, Tanja and Helbach, Jasmin and Hoffmann, Amrei and Mathes, Tim and Pieper, Dawid},
  title = {Nearly 80 Systematic Reviews Were Published Each Day: Observational Study on Trends in Epidemiology and Reporting over the Years 2000-2019},
  journal = {Journal of Clinical Epidemiology},
  volume = {138},
  pages = {1--11},
  year = {2021},
  doi = {10.1016/j.jclinepi.2021.05.022},
  url = {https://doi.org/10.1016/j.jclinepi.2021.05.022}
}

@article{page2021prisma,
  author = {Page, Matthew J. and McKenzie, Joanne E. and Bossuyt, Patrick M. and Boutron, Isabelle and Hoffmann, Tammy C. and Mulrow, Cynthia D. and Shamseer, Larissa and Tetzlaff, Jennifer M. and Akl, Elie A. and Brennan, Sue E. and Chou, Roger and Glanville, Julie and Grimshaw, Jeremy M. and Hrobjartsson, Asbjorn and Lalu, Manoj M. and Li, Tianjing and Loder, Elizabeth W. and Mayo-Wilson, Evan and McDonald, Steve and McGuinness, Luke A. and Stewart, Lesley A. and Thomas, James and Tricco, Andrea C. and Welch, Vivian A. and Whiting, Penny and Moher, David},
  title = {The PRISMA 2020 Statement: An Updated Guideline for Reporting Systematic Reviews},
  journal = {BMJ},
  volume = {372}, pages = {n71},
  year = {2021},
  doi = {10.1136/bmj.n71},
  url = {https://doi.org/10.1136/bmj.n71}
}

@article{rethlefsen2021prismaS,
  author = {Rethlefsen, Melissa L. and Kirtley, Shona and Waffenschmidt, Siw and Ayala, Ana Patricia and Moher, David and Page, Matthew J. and Koffel, Jonathan B. and Blunt, Heather and Brigham, Tara and Chang, Steven and Clark, Justin and Conway, Aislinn and Couban, Rachel and de Kock, Shelley and Farrah, Kelly and Fehrmann, Paul and Foster, Margaret and Fowler, Susan A. and Glanville, Julie and Harris, Elizabeth and Hoffecker, Lilian and Isojarvi, Jaana and Kaunelis, David and Ket, Hans and Levay, Paul and Lyon, Jennifer and McGowan, Jessie and Murad, M. Hassan and Nicholson, Joey and Pannabecker, Virginia and Paynter, Robin and Pinotti, Rachel and Ross-White, Amanda and Sampson, Margaret and Shields, Tracy and Stevens, Adrienne and Sutton, Anthea and Weinfurter, Elizabeth and Wright, Kath and Young, Sarah},
  title = {PRISMA-S: An Extension to the PRISMA Statement for Reporting Literature Searches in Systematic Reviews},
  journal = {Systematic Reviews},
  volume = {10},
  number = {1},
  year = {2021},
  pages = {39}, doi = {10.1186/s13643-020-01542-z},
  url = {https://doi.org/10.1186/s13643-020-01542-z}
}

@article{cacciamani2023prismaAi,
  author = {Cacciamani, Giovanni E. and Chu, Timothy N. and Sanford, Daniel I. and Abreu, Andre and Duddalwar, Vinay and Oberai, Assad and Kuo, C.-C. Jay and Liu, Xiaoxuan and Denniston, Alastair K. and Vasey, Baptiste and McCulloch, Peter and Wolff, Robert F. and Mallett, Sue and Mongan, John and Kahn, Charles E. and Sounderajah, Viknesh and Darzi, Ara and Dahm, Philipp and Moons, Karel G. M. and Topol, Eric and Collins, Gary S. and Moher, David and Gill, Inderbir S. and Hung, Andrew J.},
  title = {PRISMA AI reporting guidelines for systematic reviews and meta-analyses on AI in healthcare},
  journal = {Nature Medicine},
  volume = {29},
  number = {1},
  pages = {14--15},
  year = {2023},
  doi = {10.1038/s41591-022-02139-w},
  url = {https://doi.org/10.1038/s41591-022-02139-w}
}

@article{holst2025prismaTraice,
  author = {Holst, Dirk and Moenck, Keno and Koch, Julian and Schmedemann, Ole and Schuppstuhl, Thorsten},
  title = {Transparent Reporting of AI in Systematic Literature Reviews: Development of the PRISMA-trAIce Checklist},
  journal = {JMIR AI},
  volume = {4},
  pages = {e80247},
  year = {2025},
  doi = {10.2196/80247},
  url = {https://ai.jmir.org/2025/1/e80247}
}

@misc{susnjak2023prismaDfllm,
  author = {Susnjak, Teo},
  title = {PRISMA-DFLLM: An Extension of PRISMA for Systematic Literature Reviews using Domain-specific Finetuned Large Language Models},
  year = {2023},
  howpublished = {arXiv:2306.14905},
  doi = {10.48550/arXiv.2306.14905},
  url = {https://arxiv.org/abs/2306.14905}
}

@misc{shailendra2026lPrisma,
  author = {Shailendra, Samar and Kadel, Rajan and Sharma, Aakanksha and Tahidul, Islam Mohammad and Saxena, Urvashi Rahul},
  title = {L-PRISMA: An Extension of PRISMA in the Era of Generative Artificial Intelligence (GenAI)},
  year = {2026},
  howpublished = {arXiv:2603.19236},
  url = {https://arxiv.org/abs/2603.19236}
}

@article{zabaleta2026scilitbench,
  title = {{SciLitBench}: Benchmark and Design Principles for {LLM}-Powered Systematic Literature Reviews},
  author = {Zabaleta, Miguel and Lin, Baihan},
  journal = {arXiv preprint arXiv:2609.05505},
  year = {2026},
  doi = {10.48550/arXiv.2609.05505},
  url = {https://arxiv.org/abs/2609.05505}
}

\clearpage
\section*{Supplementary Information}

\setcounter{subsection}{0}
\renewcommand{\thesubsection}{S\arabic{subsection}}

The supplementary material provides calculation details and supporting analyses for the empirical results, followed by the expanded \ouralgo checklist and implementation-level details.

\setcounter{figure}{0}
\renewcommand{\thefigure}{S\arabic{figure}}
\renewcommand{\theHfigure}{supp.\arabic{figure}}
\setcounter{table}{0}
\renewcommand{\thetable}{S\arabic{table}}
\renewcommand{\theHtable}{supp.\arabic{table}}

\subsection{Supplementary methods}

\noindent\textbf{Analysis denominators and multi-label fields.}
All supplementary percentages use papers as the denominator. Approach, review-stage, limitation, and model-family analyses are multi-label unless an exclusive group is named. A paper can therefore contribute to more than one approach or review-stage group, and percentages across groups need not sum to 100\%. The number of papers contributing to a cell is reported in the figure, caption, or accompanying table when needed to interpret the estimate.

\medskip\noindent\textbf{Overall impressions and high-bar concerns.}
The overall-impression analysis is restricted to papers annotated with an LLM approach. Item-level impression mappings were collapsed into four exclusive paper-level groups: positive only; positive with caveats; mixed or negative; and descriptive or unclear. Monthly values use the rolling 3-month procedure described in the Methods. The high-bar analysis uses the same paper-level impression groups and marks a paper as high-bar unmet when at least one limitation item states that the method did not meet the performance or reliability bar required for its intended use. The percentages in \panelref{fig:utility-adequacy}{b} use all LLM papers in each impression group as the denominator.

\medskip\noindent\textbf{Limitation profiles and software reporting.}
Limitation prevalence was calculated as the share of papers in a group with at least one item in the named limitation category. \suppfigref{fig:supp-limitation-profiles} reports four recurring categories used in the reporting analysis: an unmet high bar, limited validation, parameter or prompt sensitivity, and small or narrow data. The categories are not mutually exclusive. 

The software-reporting analysis includes papers published from 2023 onward. Papers were grouped by approach and review stage using the multi-label definitions in the Methods. Substantive evaluation items include performance, comparative, modification, and runtime or cost annotations; qualitative impressions, \texttt{none reported}, and \texttt{other} items are excluded. Within each approach-stage cell, we calculated the mean number of substantive evaluation items per paper and the share of papers annotated with \texttt{none reported}.

\medskip\noindent\textbf{Reported performance evidence.}
Metric names and values were parsed from performance-versus-human annotations as described in the Methods. \suppfigref{fig:supp-reported-performance} and \supptabref{tab:supp-reported-performance} include only papers that reported both precision and recall for screening or data extraction. When a paper reported more than one value for the same approach, stage, and metric, values were collapsed to the paper-level median. The large diamonds in the figure mark the median precision and recall within each approach-stage group.

\subsection{Supplementary analyses}

\noindent\textbf{Positive assessments and unmet reliability requirements.}
Overall impressions of LLM-assisted review automation are frequently positive, while limitation reporting gives a more qualified view of what the methods can do. Among the 118 LLM papers with positive-only overall evaluations, 52\% also report at least one high-bar-unmet concern. The corresponding share rises to 67\% among the 57 papers with positive evaluations and caveats and to 79\% among the 38 papers with mixed or negative evaluations. Among the 35 descriptive or unclear papers, the share is 14\% (\panelref{fig:utility-adequacy}{b}). The rolling analysis shows positive assessments throughout the observed LLM period, alongside continuing caveated and mixed assessments (\panelref{fig:utility-adequacy}{a}). Usefulness and adequacy therefore emerge as separate judgments in the literature: authors can find an LLM workflow promising while also reporting that it remains below the reliability required for its intended review role.

\begin{figure}[!p]
\centering

\begin{minipage}[t]{0.49\textwidth}
\vspace{0pt}
\centering
\textbf{a) By approach}\par\vspace{0.25em}
\includegraphics[width=\linewidth]{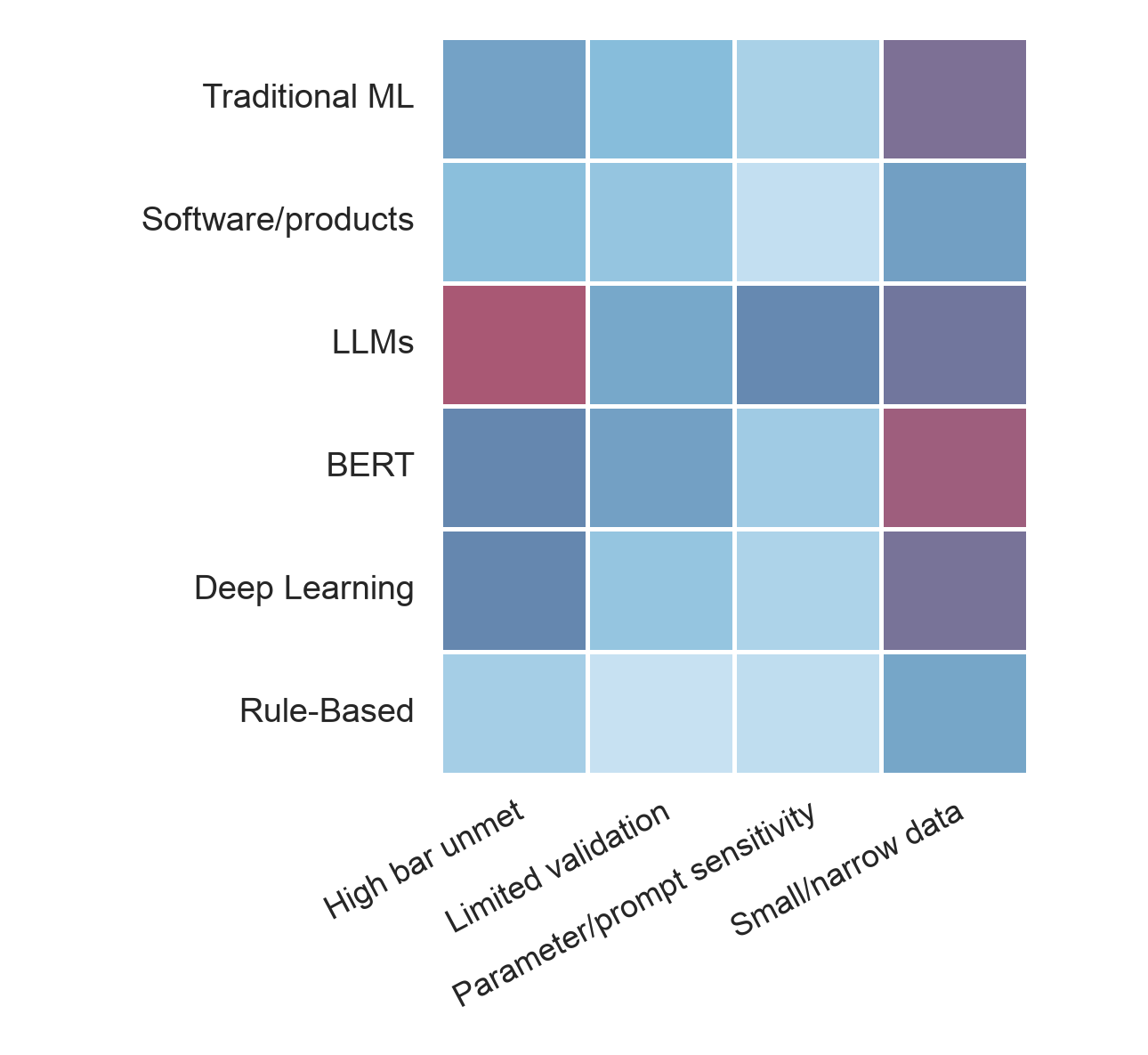}
\end{minipage}
\hfill
\begin{minipage}[t]{0.49\textwidth}
\vspace{0pt}
\centering
\textbf{b) By high-level domain}\par\vspace{0.25em}
\includegraphics[width=\linewidth]{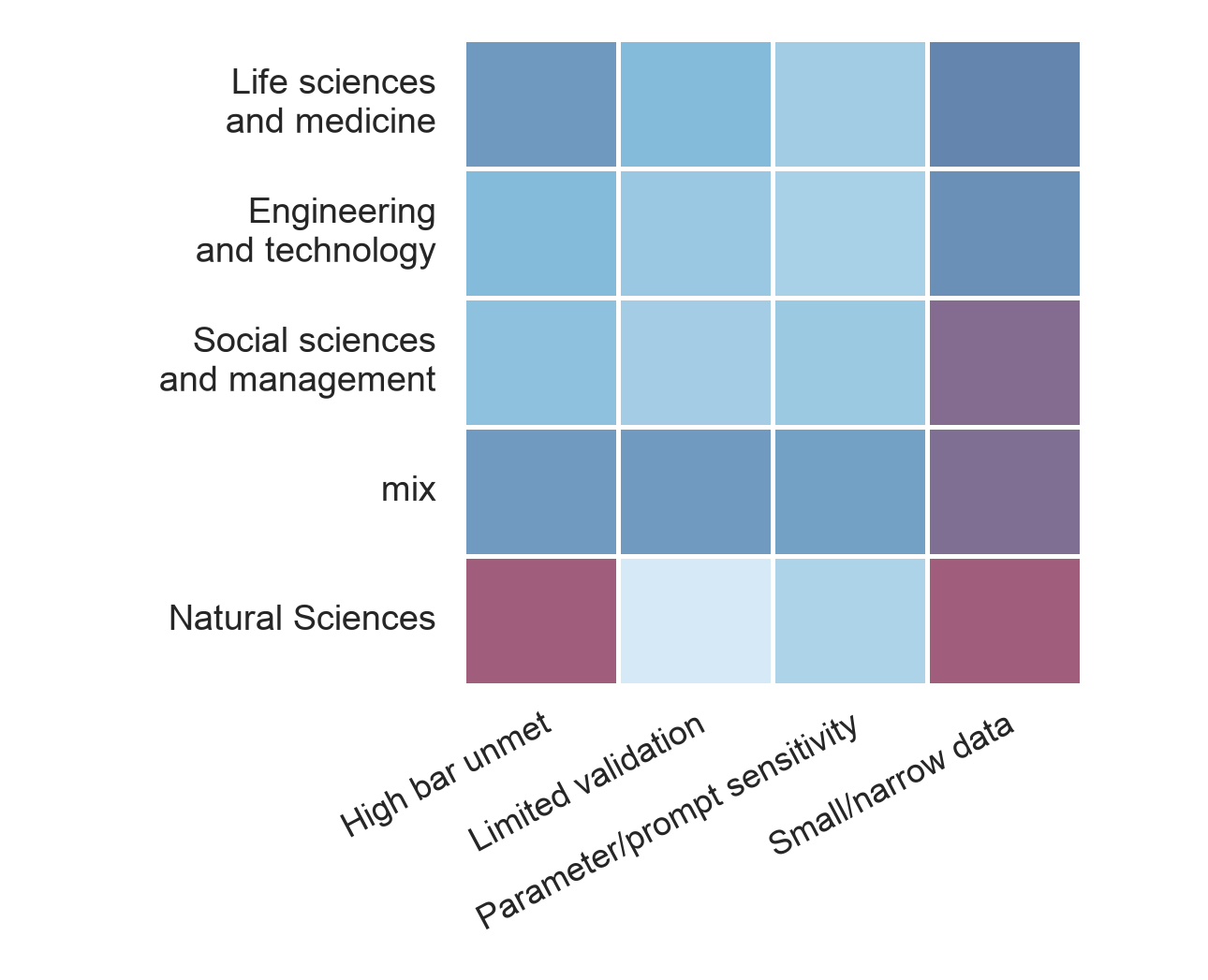}
\end{minipage}

\vspace{0.6em}

\begin{minipage}[t]{0.49\textwidth}
\vspace{0pt}
\centering
\textbf{c) By BERT family}\par\vspace{0.25em}
\includegraphics[width=\linewidth]{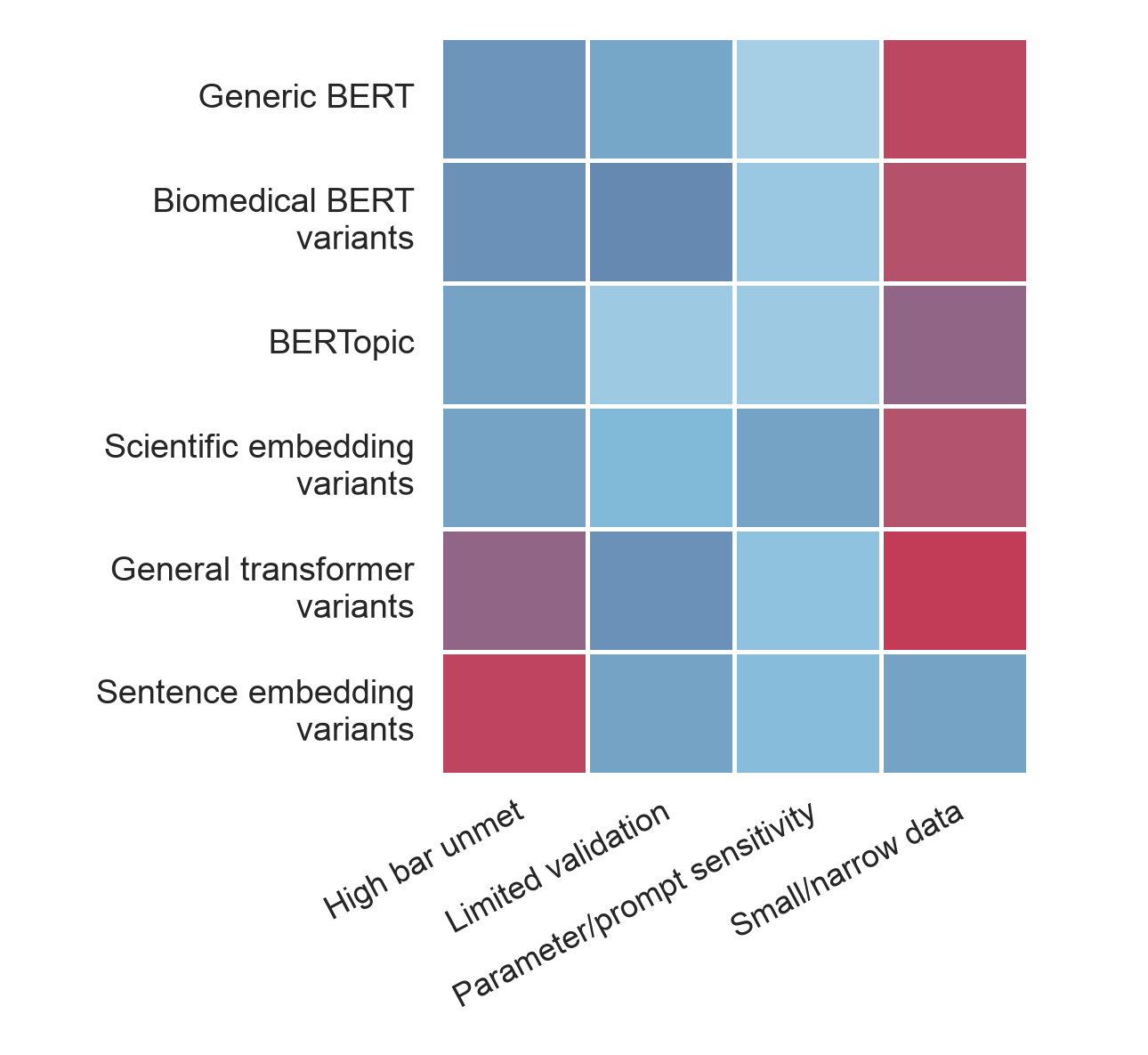}
\end{minipage}
\hfill
\begin{minipage}[t]{0.49\textwidth}
\vspace{0pt}
\centering
\textbf{d) By LLM family}\par\vspace{0.25em}
\includegraphics[width=\linewidth]{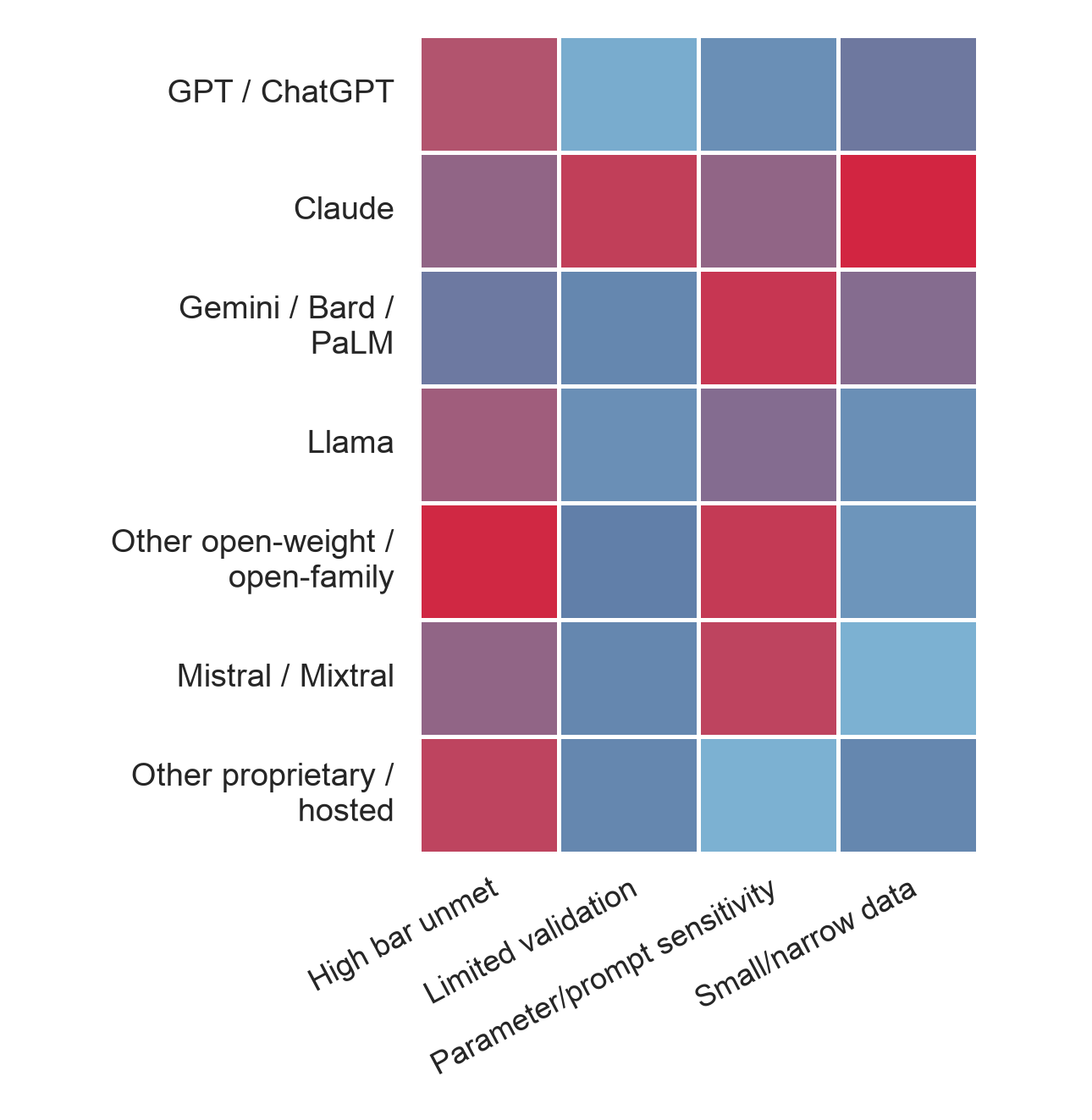}
\end{minipage}

\vspace{0.3em}

\includegraphics[width=0.52\textwidth]{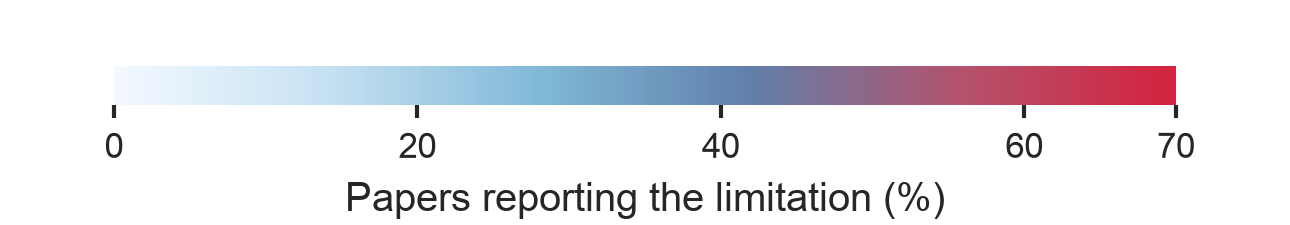}

\caption{\textbf{Recurring limitations by approach, domain, and model family.} Cells show the percentage of papers within each row group that report at least one item assigned to the corresponding limitation category. Panels group papers by \textbf{a)} approach, \textbf{b)} high-level domain, \textbf{c)} BERT family, and \textbf{d)} LLM family. All panels use the same 0--70\% color scale, with higher prevalence shown in red. Limitation categories are not mutually exclusive, so a paper can contribute to multiple columns. Papers can also contribute to multiple approach or model-family groups; those panels therefore show within-group prevalence rather than mutually exclusive partitions of the corpus. Model-family estimates should be interpreted alongside group size because several families contain few papers.}

\label{fig:supp-limitation-profiles}
\end{figure}

\medskip\noindent\textbf{Limitation profiles vary across methods.}
The kinds of limitations authors report also differ across automation approaches. Among LLM papers, 54\% report an unmet high bar, 40\% report parameter or prompt sensitivity, 32\% report limited validation, and 45\% report small or narrow data. Software/product papers report the same categories less often: 26\%, 15\%, 24\%, and 34\%, respectively. Earlier approaches show a different profile. Small or narrow data is reported by 47\% of traditional-ML papers and 52\% of BERT papers, while high-bar-unmet concerns appear in 33\% and 40\% (\suppfigref{fig:supp-limitation-profiles}). The reported limitation profiles therefore differ by method. Prompt sensitivity appears especially often in LLM papers, whereas data scope remains prominent across both earlier and current approaches.

\begin{figure}[!htbp]
\centering
\begin{minipage}[t]{0.49\textwidth}
\textbf{a) Mean evaluation items}\par
\vspace{2pt}
\centering
\includegraphics[width=\linewidth]{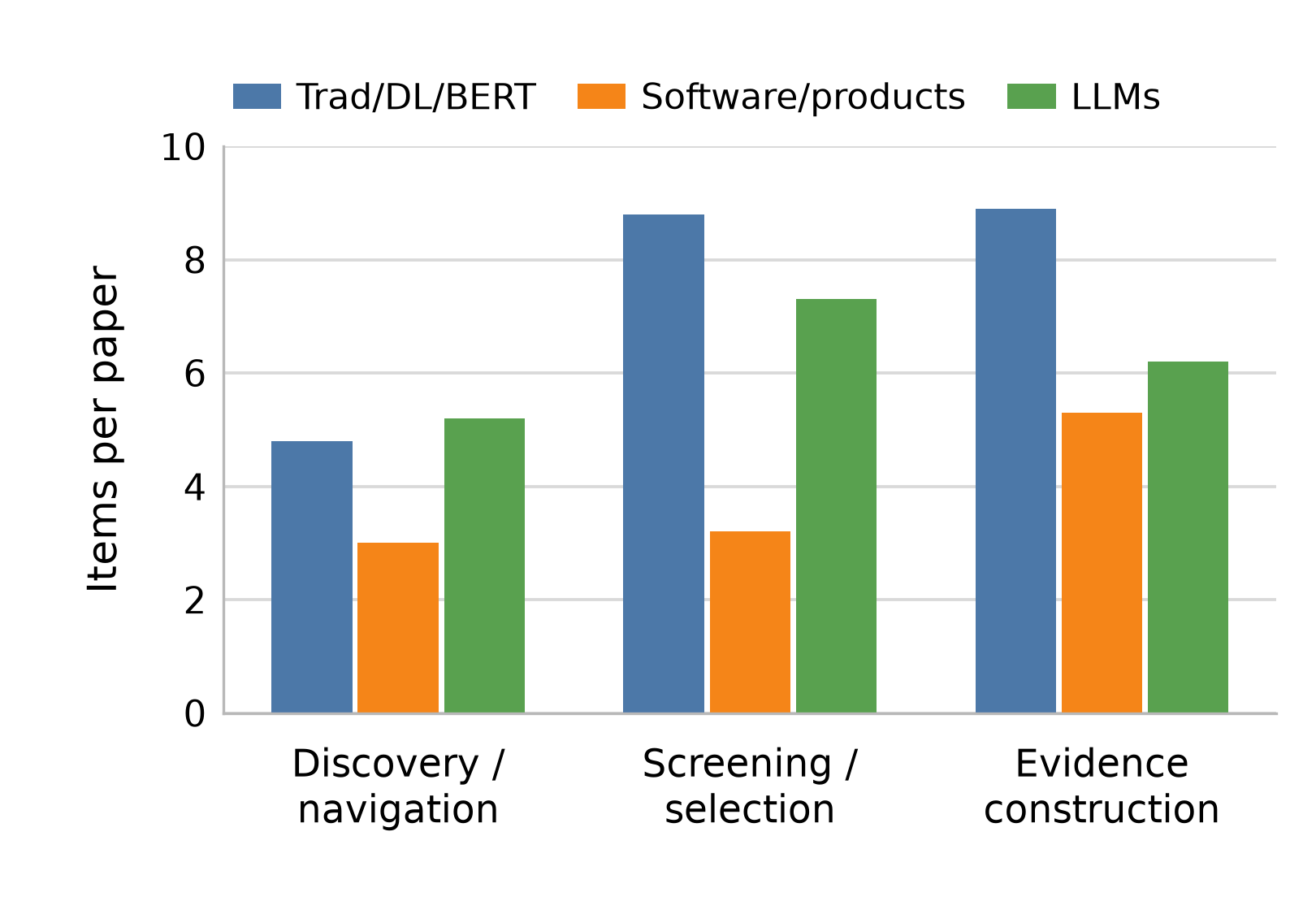}
\end{minipage}
\hfill
\begin{minipage}[t]{0.49\textwidth}
\textbf{b) No evaluation reported}\par
\vspace{2pt}
\centering
\includegraphics[width=\linewidth]{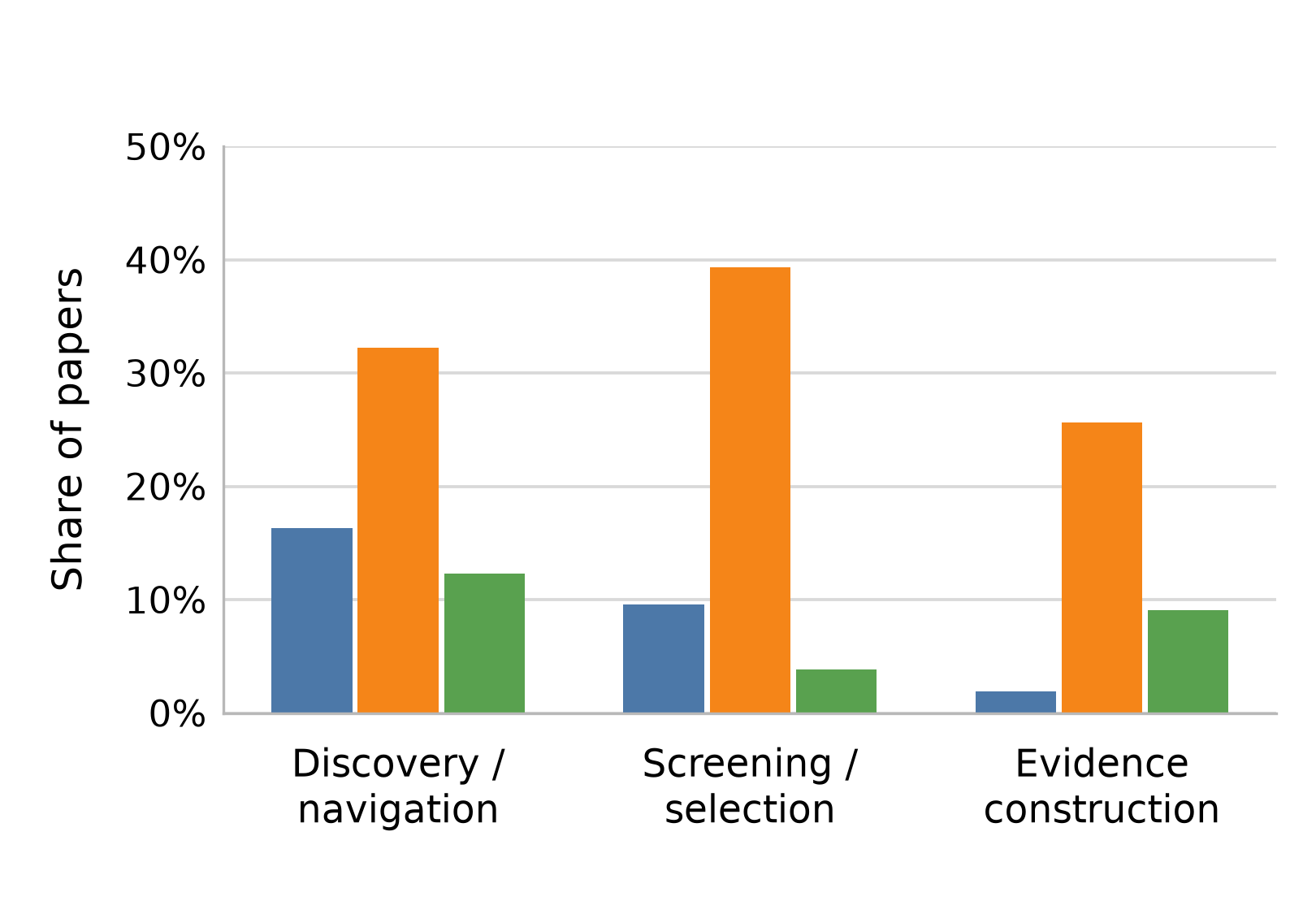}
\end{minipage}

\caption{\textbf{Evaluation reporting by approach and review stage since 2023.} Bars summarize papers published from 2023 onward within each approach--stage combination. \textbf{a)} Mean number of substantive evaluation items reported per paper. \textbf{b)} Percentage of papers reporting no evaluation. Approach and review-stage labels are multi-label, so papers can contribute to more than one group. Within each review-stage group, software/product papers report fewer evaluation items and are more likely to report no evaluation than traditional ML, deep-learning, or BERT papers and LLM papers. The aggregate difference is therefore not explained solely by differences in review-stage composition.}
\label{fig:supp-software-reporting}
\end{figure}

\medskip\noindent\textbf{The software reporting gap persists across review stages.}
The lower reporting observed for software/product papers appears within each review-stage group. In discovery and navigation, software/product papers report 3.0 substantive evaluation items on average, compared with 5.2 for LLM papers and 4.8 for traditional, deep-learning, or BERT papers. Their no-evaluation rate is also higher at 32.5\%, compared with 12.3\% and 16.3\%. The difference is largest in screening and selection: software/product papers average 3.2 evaluation items and 39.3\% report no evaluation, whereas LLM papers average 7.3 items with a 3.8\% no-evaluation rate. Evidence construction shows the same direction, with means of 5.3 for software/products, 6.2 for LLMs, and 9.0 for earlier custom approaches (\suppfigref{fig:supp-software-reporting}). Stage-stratified comparisons preserve the software-reporting gap, showing that the aggregate difference is not explained solely by review-stage composition.

\begin{figure}[!htbp]
\centering

\begin{minipage}[t]{0.49\textwidth}
\textbf{a) Screening and selection}\par
\vspace{2pt}
\centering
\includegraphics[width=\linewidth]
{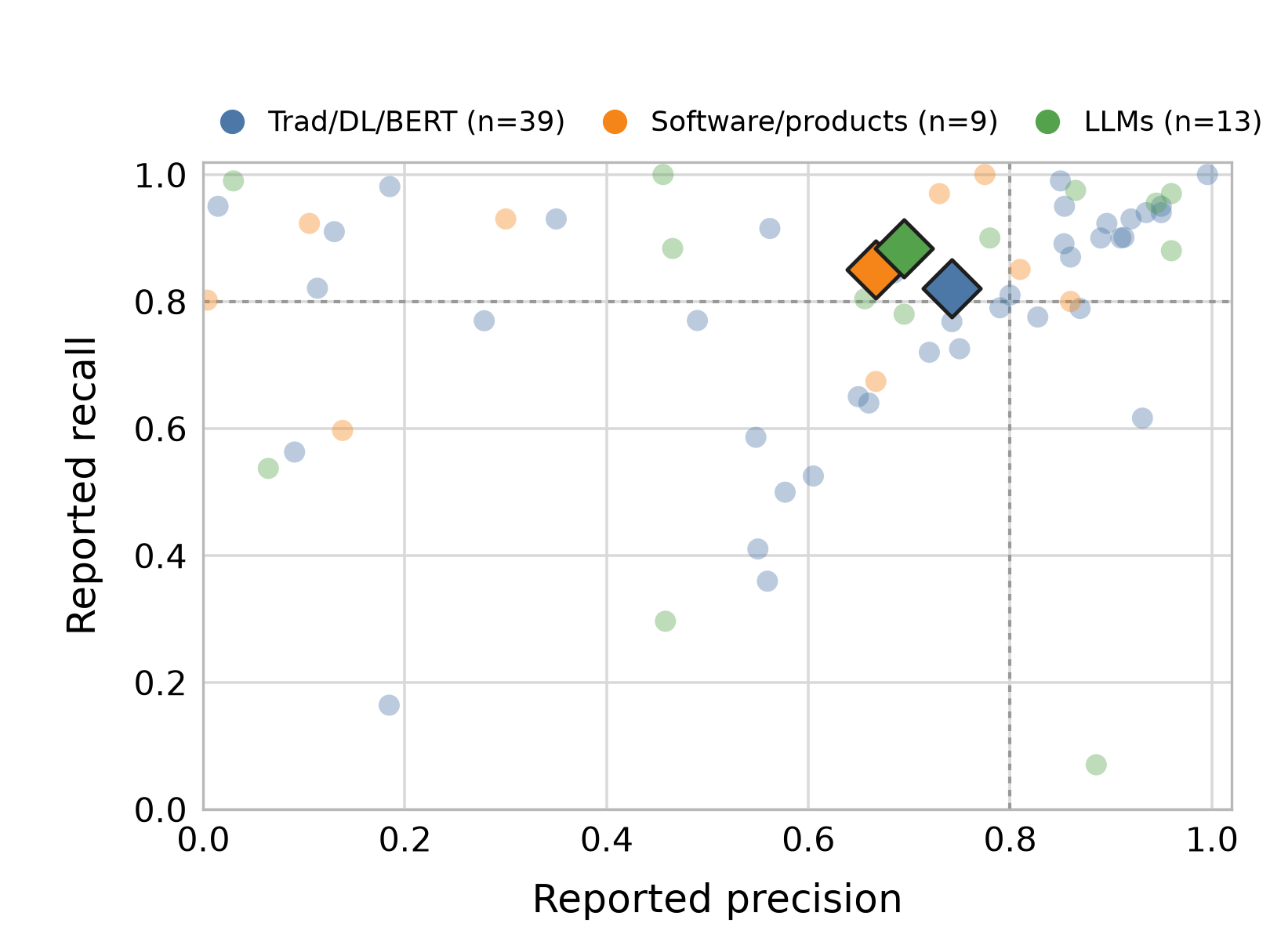}
\end{minipage}
\hfill
\begin{minipage}[t]{0.49\textwidth}
\textbf{b) Data extraction}\par
\vspace{2pt}
\centering
\includegraphics[width=\linewidth]
{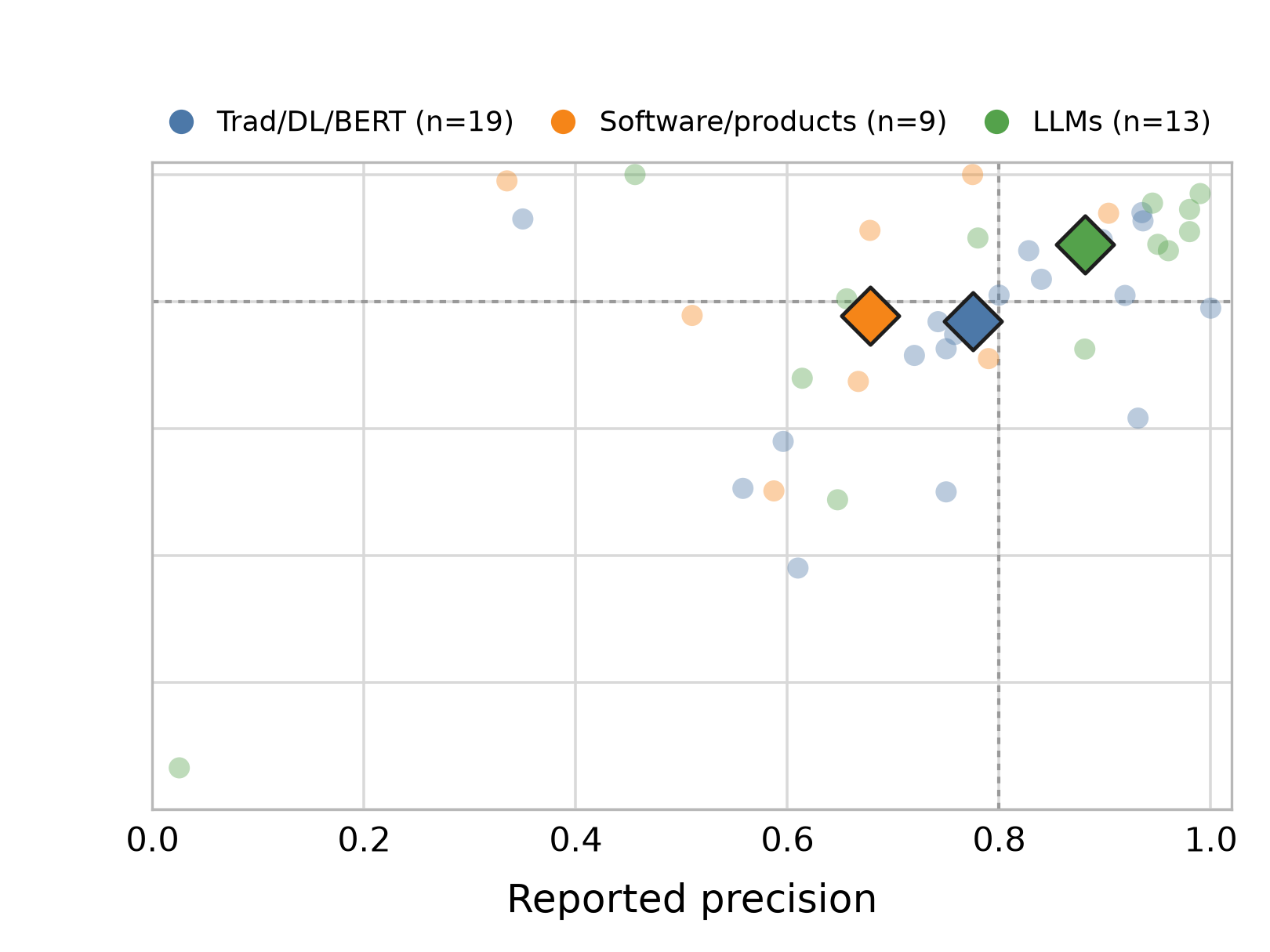}
\end{minipage}

\caption{\textbf{Reported precision and recall for screening and data extraction.} \textbf{a)} Screening and selection. \textbf{b)} Data extraction. Each small point represents one paper that reported both precision and recall, colored by approach group; legend values give the number of papers. Large diamonds mark the within-group medians for the two metrics. The dotted vertical and horizontal lines mark precision and recall of 0.8, respectively. Values come from heterogeneous studies using different datasets, tasks, thresholds, and evaluation designs.}
\label{fig:supp-reported-performance}
\end{figure}

\medskip\noindent\textbf{Reported performance evidence is heterogeneous.}
Papers reporting both precision and recall show substantial variation within every approach group. For screening and selection, median recall is 0.821 for custom ML, deep-learning, BERT, or rule-based methods, 0.884 for LLMs, and 0.851 for software/products; median precision is 0.742, 0.695, and 0.667, respectively. For data extraction, median recall is 0.769, 0.890, and 0.778, while median precision is 0.776, 0.881, and 0.678 (\suppfigref{fig:supp-reported-performance} and \supptabref{tab:supp-reported-performance}). The individual paper values span much wider ranges than the group medians suggest. Given this dispersion and variation in datasets, tasks, thresholds, and evaluation designs, these metrics provide a descriptive inventory of the evidence reported by the studies. The observed variation also supports reporting task-specific performance together with the evaluation context needed to interpret it.


\subsection{Supplementary Figures}

\begin{figure}[!htbp]
\centering

\begin{minipage}{0.96\textwidth}
\textbf{a) BERT-family composition}\par
\vspace{2pt}
\centering
\includegraphics[width=\linewidth]{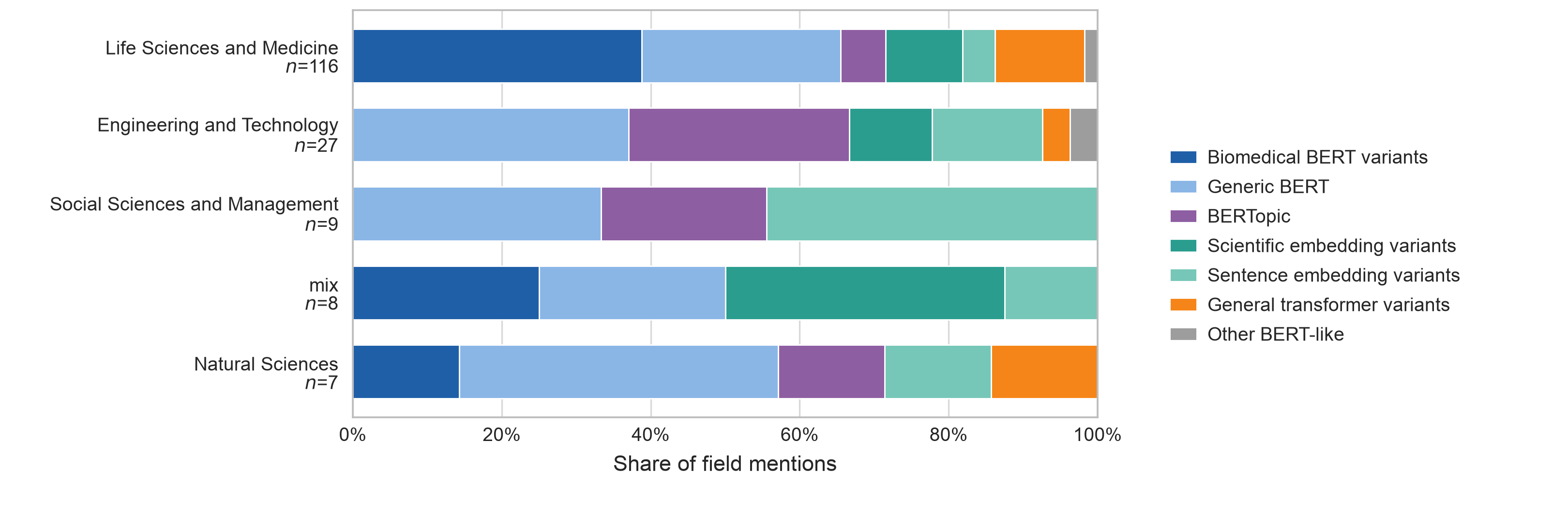}
\end{minipage}

\vspace{6pt}

\begin{minipage}{0.96\textwidth}
\textbf{b) LLM model-family composition}\par
\vspace{2pt}
\centering
\includegraphics[width=\linewidth]{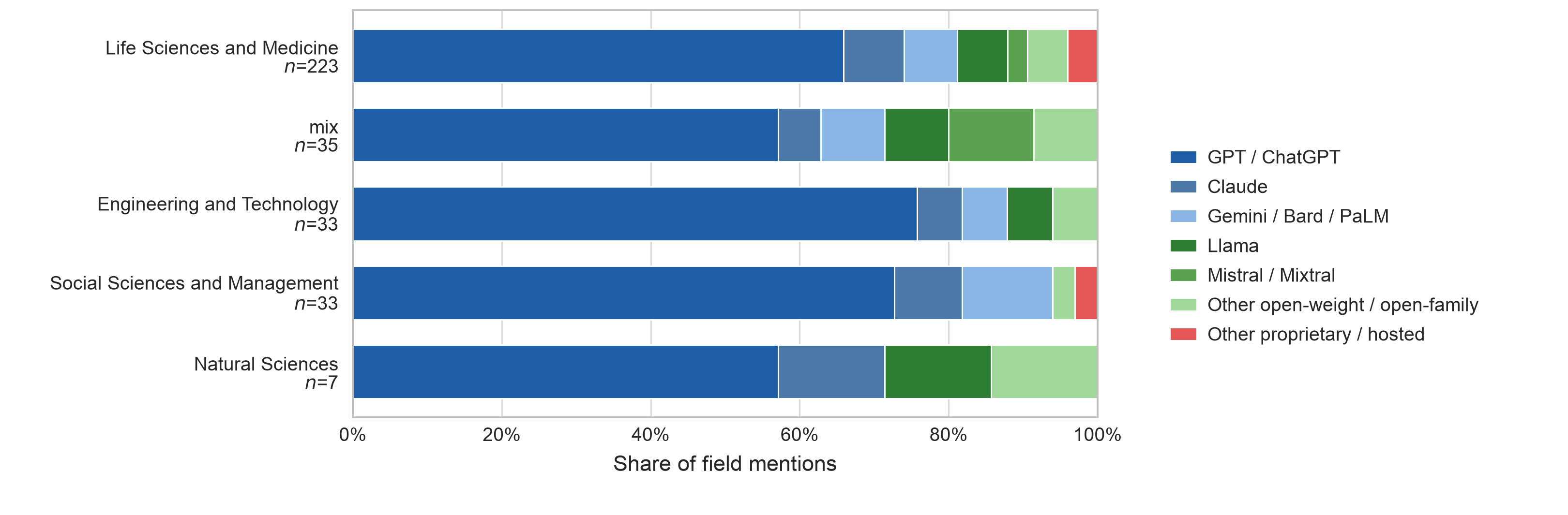}
\end{minipage}

\caption{\textbf{Model-family use by field.} Bars show within-field distributions of model-family mentions. Because a paper may mention more than one family, displayed $n$ values are paper--family mentions rather than unique papers. Both panels include papers published from 2015 onward; arts and humanities is omitted because it contains only two papers. \textbf{a)} BERT-family composition. Biomedical variants are most prominent in life sciences and medicine, whereas other fields contain a broader mixture of generic BERT, BERTopic, embedding, and transformer variants. \textbf{b)} LLM-family composition among papers with an identifiable LLM family. GPT/ChatGPT-family systems dominate across fields; Claude, Gemini/Bard/PaLM, Llama, Mistral/Mixtral, and other families appear less often. The panels describe reported model use and do not compare model performance.}
\label{fig:supp-model-family}
\end{figure}

\begin{figure}[!tbp]
\centering

\begin{minipage}[t]{0.49\textwidth}
\raggedright
\textbf{a) Approach adoption over time}\par\smallskip
\includegraphics[width=\linewidth]
{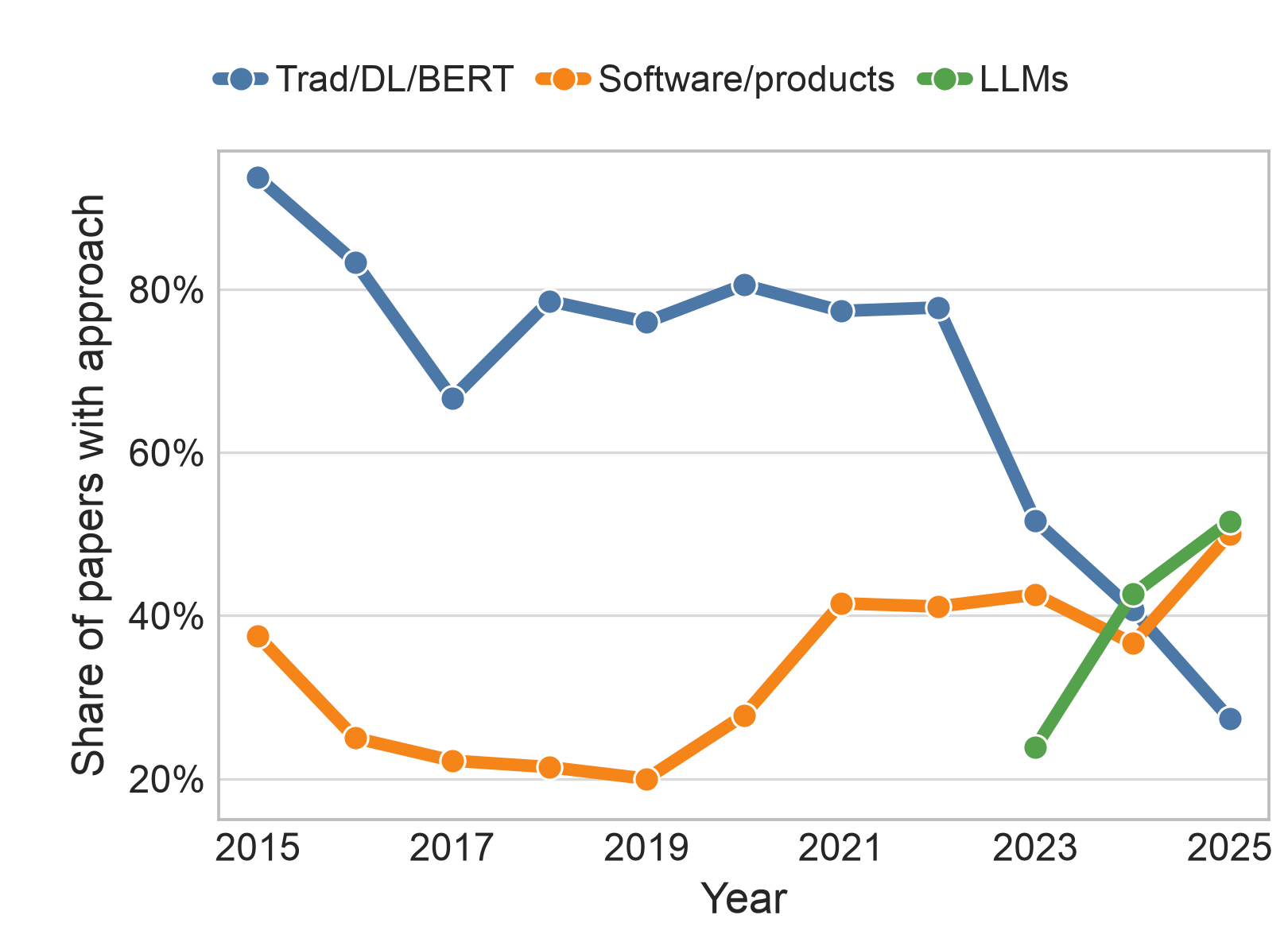}
\end{minipage}
\hfill
\begin{minipage}[t]{0.49\textwidth}
\raggedright
\textbf{b) Reporting richness over time}\par\smallskip
\includegraphics[width=\linewidth]
{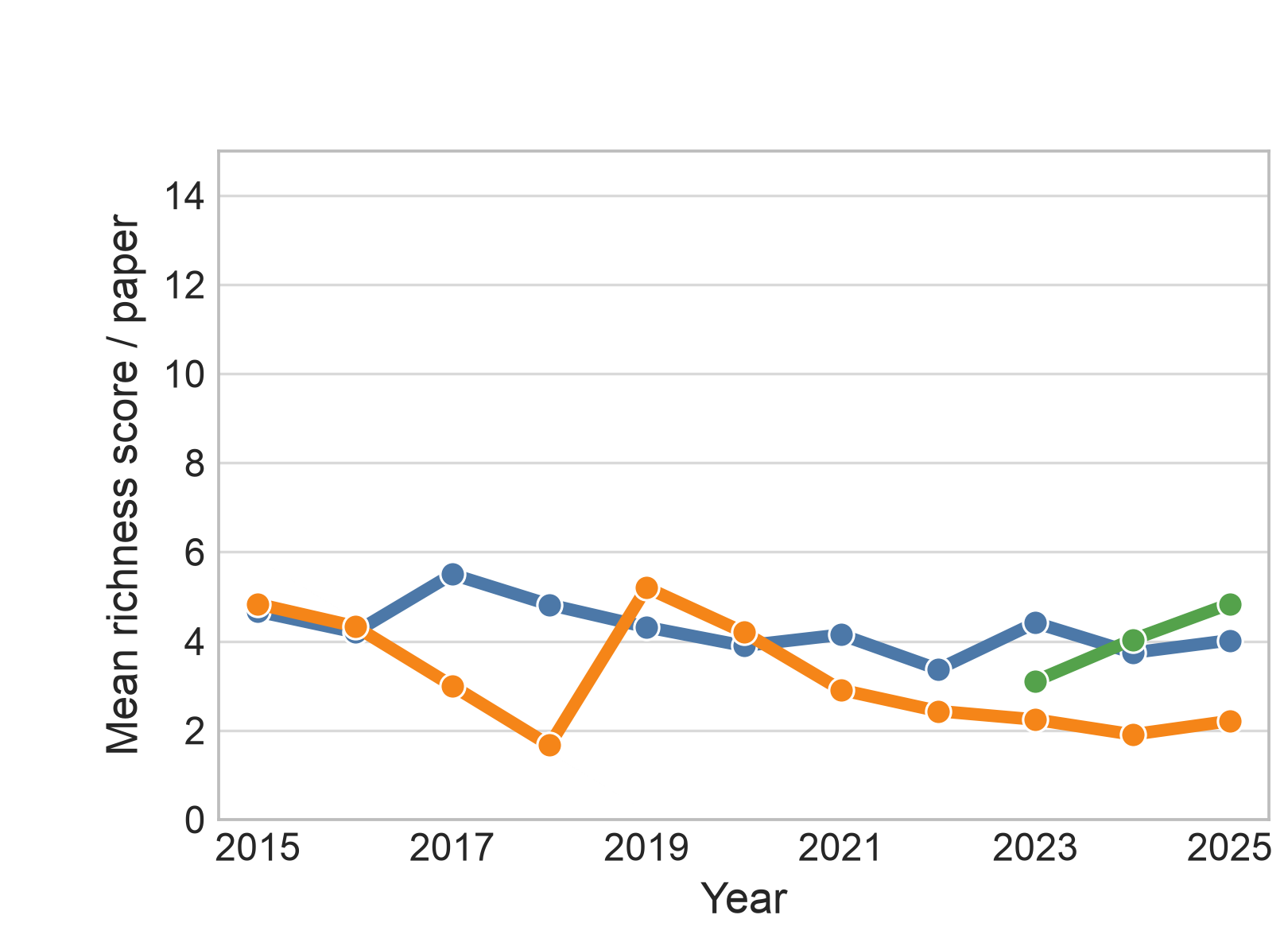}
\end{minipage}

\caption{\textbf{Approach prevalence and evaluation reporting over time.}
\textbf{a)} Annual share of papers carrying each approach label. Approach categories are multi-label, so shares may sum to more than 100\% across groups.
\textbf{b)} Mean paper-level reporting-richness score by approach and publication year. The 0--15 index summarizes five dimensions of reported evaluation and limitations and should not be interpreted as study quality. LLM papers show increasing reporting coverage after 2023, whereas software/product papers remain common but report less paper-level evaluation and limitation detail. Values for 2025 include January through June only.}

\label{fig:supp-evaluation-reporting}
\end{figure}

\begin{figure}[!tbp]
\centering
\includegraphics[width=.95\textwidth]{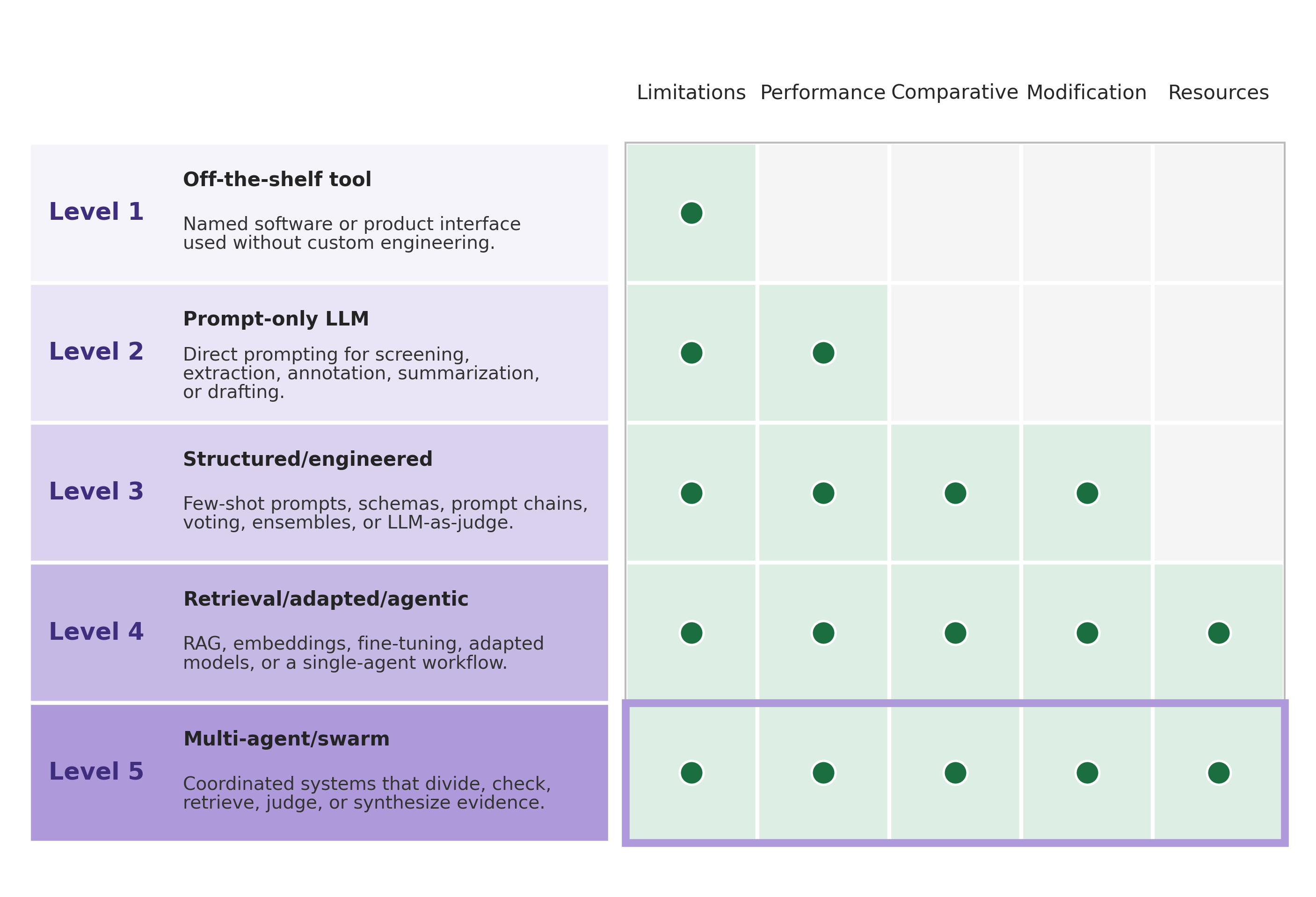}
\vspace{-1em}
\caption{\textbf{\ouralgo evaluation and limitation reporting map.} Rows define five proposed levels of workflow complexity, and marked cells indicate the minimum reporting dimensions associated with each level. Expectations are cumulative: all workflows should report limitations or failure modes; prompt-only workflows add task-specific performance evidence; structured or engineered workflows add comparative and modification evidence; and retrieval, adapted, agentic, and multi-agent workflows cover all five dimensions. 
The levels reflect current review-automation practice and should be revised as the field changes. 
At any level, stronger task-specific validation is expected when AI outputs can alter retrieval, inclusion/exclusion, extracted evidence, appraisal, synthesis, or conclusions; when human verification is incomplete; when the system is opaque or mutable; or when errors are difficult to reconstruct or reverse.}

\label{fig:prisma-llm-reporting-map}
\end{figure}

\clearpage
\subsection{Supplementary Tables}

\begin{table}[!htbp]
\centering
\footnotesize
\setlength{\tabcolsep}{4pt}
\renewcommand{\arraystretch}{1.08}
\begin{minipage}[t]{0.43\textwidth}
\vspace{0pt}
\begin{tabular}{@{}L{0.56\linewidth}rr@{}}
\toprule
Field or domain & n & Share \\
\midrule
\multicolumn{3}{@{}l}{\textit{Record fields}} \\
year & 888 &  \\
domain & 1776 &  \\
review stage & 1350 &  \\
approach & 1254 &  \\
evaluation results & 6562 &  \\
limitations & 2896 &  \\
\textbf{total} & \textbf{14726} &  \\
\addlinespace[0.45em]
\multicolumn{3}{@{}l}{\textit{Domains}} \\
Life sciences/medicine & 548 & 61.7\% \\
Engineering/technology & 140 & 15.8\% \\
Social sciences/management & 115 & 13.0\% \\
Mixed domain & 62 & 7.0\% \\
Natural sciences & 21 & 2.4\% \\
Arts/humanities & 2 & 0.2\% \\
\bottomrule
\end{tabular}
\end{minipage}
\hfill
\begin{minipage}[t]{0.53\textwidth}
\vspace{0pt}
\begin{tabular}{@{}L{0.66\linewidth}rr@{}}
\toprule
Annotation category & Items & Papers \\
\midrule
\multicolumn{3}{@{}l}{\textit{Evaluation results}} \\
performance vs human annotation & 1788 & 395 \\
overall sentiment / qualitative impression & 1778 & 663 \\
comparative results & 1149 & 386 \\
feature or modification effects & 959 & 353 \\
runtime and cost & 722 & 320 \\
none reported & 163 & 163 \\
\addlinespace[0.45em]
\multicolumn{3}{@{}l}{\textit{Limitations}} \\
small data, narrow scope, or limited data availability & 698 & 373 \\
high bar needed but not achieved & 642 & 295 \\
limited validation or benchmarking of AI outputs & 445 & 229 \\
performance sensitive to parameters or prompts & 393 & 196 \\
high human workload demands & 256 & 146 \\
subjectivity and labeling bias & 199 & 118 \\
none reported & 238 & 238 \\
\bottomrule
\end{tabular}
\end{minipage}
\par\vspace{7pt}
\caption{\textbf{Corpus composition and annotation inventory.} The left block reports item counts across the six annotated fields and the distribution of 888 papers across mutually exclusive high-level domains. The right block reports item counts for recurring evaluation-result and limitation categories and the number of papers contributing at least one item to each category. Papers can contribute multiple items and can appear in more than one category, so category-level paper counts are not additive.}
\label{tab:supp-corpus-inventory}
\end{table}

\begin{table}[!htbp]
\centering

{\footnotesize
\setlength{\tabcolsep}{4pt}
\renewcommand{\arraystretch}{1.08}
\begin{tabular}{@{}L{0.21\textwidth}L{0.53\textwidth}L{0.22\textwidth}@{}}
\toprule
Basis & Contribution to the checklist & Related items \\
\midrule

SciLitBench reporting schema
& The annotation schema identifies five recurring reporting dimensions: performance, comparisons, modifications, resources or feasibility, and limitations. These dimensions define reporting richness and the graded evaluation requirements.
& LLM-M11, LLM-R2, LLM-R4, LLM-D1; \supptabref{tab:prisma-llm-complexity}. \\

SciLitBench complexity analysis
& Reporting richness increases with workflow complexity in the observed corpus, motivating requirements that accumulate across complexity levels.
& LLM-M3, LLM-M11, LLM-R2; \supptabref{tab:prisma-llm-complexity}. \\

SciLitBench screening workflow
& Screening automation requires reporting how outputs determine inclusion, exclusion, prioritization, or progression to full-text review, together with audit and adjudication procedures.
& LLM-M7, LLM-M10, LLM-R1, LLM-R3. \\

SciLitBench extraction and harmonization workflow
& Structured extraction requires schemas, source evidence, edge-case rules, human calibration, and adjudication. Harmonization also requires recoverable final labels and decision rules.
& LLM-M5, LLM-M8, LLM-M9, LLM-M10, LLM-O1. \\

SciLitBench document processing
& Full-text automation can depend on PDF parsing, OCR, table and section handling, context limits, chunking, truncation, and access to supplementary files.
& LLM-M4. \\

SciLitBench workflow development
& Prompt, schema, and workflow revisions expose failed approaches and the design choices needed to obtain usable outputs.
& LLM-M6, LLM-R3, LLM-D2. \\

PRISMA 2020 and PRISMA-S
\citep{page2021prisma,rethlefsen2021prismaS}
& Established guidance requires reporting review methods and technically consequential search decisions at the protocol, method, result, and availability levels.
& LLM-T1, LLM-A1, LLM-I1, LLM-M1, LLM-M12, LLM-O1, LLM-O2. \\

Related AI-reporting proposals
\citep{cacciamani2023prismaAi,holst2025prismaTraice,susnjak2023prismaDfllm,shailendra2026lPrisma}
& Related proposals cover system identification, prompts and model development, human interaction, performance evaluation, reproducibility, limitations, and implications of AI use.
& LLM-M2, LLM-M6, LLM-M10--M12, LLM-R2--R4, LLM-D1, LLM-D2, LLM-O1. \\

\bottomrule
\end{tabular}
}

\par\vspace{7pt}

\caption{\textbf{Evidence and guidance informing the \ouralgo checklist.}
Each row maps checklist items to findings from the SciLitBench corpus,
methodological lessons from building the SciLitBench workflow, or established
and adjacent reporting guidance. The table distinguishes empirical,
workflow-derived, and prior-guidance rationales.}
\label{tab:supp-evidence}
\end{table}

\begin{table*}[!htbp]
\centering
\footnotesize
\setlength{\tabcolsep}{4pt}
\renewcommand{\arraystretch}{1.12}
\begin{tabular}{@{}L{0.19\textwidth}L{0.23\textwidth}C{0.07\textwidth}C{0.18\textwidth}C{0.18\textwidth}@{}}
\toprule
Review task & Approach group & Papers & Precision, median [IQR] & Recall, median [IQR] \\
\midrule
Screening/selection & Custom ML/DL/BERT/rules & 39 & 0.742 [0.549--0.880] & 0.821 [0.685--0.927] \\
Screening/selection & LLMs & 13 & 0.695 [0.458--0.886] & 0.884 [0.780--0.970] \\
Screening/selection & Software/products & 9 & 0.667 [0.138--0.775] & 0.851 [0.800--0.930] \\
\addlinespace[0.35em]
Data extraction & Custom ML/DL/BERT/rules & 19 & 0.776 [0.731--0.908] & 0.769 [0.666--0.857] \\
Data extraction & LLMs & 13 & 0.881 [0.647--0.960] & 0.890 [0.725--0.945] \\
Data extraction & Software/products & 9 & 0.678 [0.587--0.783] & 0.778 [0.710--0.939] \\
\bottomrule
\end{tabular}
\par\vspace{7pt}
\caption{\textbf{Reported precision and recall for screening and data extraction.} For each review-task and approach-group combination, the table gives the number of papers reporting both metrics and the paper-level median and interquartile range for precision and recall. Source studies differ in datasets, tasks, decision thresholds, and validation designs.}
\label{tab:supp-reported-performance}
\end{table*}

\clearpage
\begingroup
\footnotesize
\setlength{\LTcapwidth}{0.94\textwidth} \begin{longtable}{@{}L{0.25\textwidth}L{0.69\textwidth}@{}}
\toprule
Structural element and item ID & Checklist item \\
\midrule
\endfirsthead
\toprule
Structural element and item ID & Checklist item \\
\midrule
\endhead
\endfoot

\multicolumn{2}{@{}l}{\textit{Title}} \\
LLM-T1 - Title & If LLMs or LLM-enabled software played a substantive role in the review workflow, consider indicating this in the title or subtitle, especially when they affected screening, extraction, synthesis, or other review decisions. \\
\addlinespace[0.35em]

\multicolumn{2}{@{}l}{\textit{Abstract}} \\
LLM-A1 - Abstract & Briefly summarize the LLM or software systems used, the review stages at which they were applied, and whether their outputs informed review decisions, extracted evidence, analysis, synthesis, or writing only. \\
\addlinespace[0.35em]

\multicolumn{2}{@{}l}{\textit{Introduction}} \\
LLM-I1 - Rationale & State the rationale for using LLMs or LLM-enabled software in the review, including the problem they were intended to address, such as screening burden, full-text retrieval, extraction burden, annotation, synthesis, quality control, or feasibility. \\
\addlinespace[0.35em]

\multicolumn{2}{@{}l}{\textit{Methods}} \\
LLM-M1 - Protocol and deviations & State whether LLM or software-mediated procedures were specified in the protocol. Report any additions, removals, or changes to LLM use after piloting, screening, extraction, annotation, or synthesis began. \\
LLM-M2 - System identification and access & For each LLM, AI tool, or LLM-enabled software product, report the name, model or version, provider, access date, interface or API, deployment mode, and whether the system was proprietary, open, local, hosted, or custom-built. \\
LLM-M3 - Review stage, task, and disclosure level & Specify the review stage and task for each system and assign a \ouralgo implementation-disclosure level using \supptabref{tab:prisma-llm-complexity}. Report whether outputs could alter search retrieval, inclusion or exclusion, extracted evidence, risk-of-bias judgments, synthesis, or conclusions; state the extent of human verification and who retained final responsibility. \\
LLM-M4 - Inputs and document processing & Describe the input data provided to the system, including records, abstracts, PDFs, full texts, tables, supplements, labels, rubrics, examples, schemas, or prior human decisions. When full texts were used, describe PDF parsing, OCR, section extraction, table handling, chunking or splitting, context-window constraints, truncation or capping rules, retrieval setup, and whether supplementary materials were included. \\
LLM-M5 - Outputs, schemas, and post-processing & Describe the outputs generated by the system, including labels, rankings, classifications, extracted fields, structured JSON, summaries, judgments, metrics, or drafted text. Report output schemas, parsing rules, confidence scores if used, automated post-processing, format repair, and how outputs were converted from free-form responses into review decisions, records, or analysis data. \\
LLM-M6 - Prompting and workflow development & Report prompts, system instructions, settings, examples, and output constraints. When prompts or workflows were revised, explain the rationale for the final version, the failure modes that motivated changes, and important alternatives or failed attempts when they affected the final method. \\
LLM-M7 - Screening and selection workflow & For title/abstract or full-text screening, report how LLM or software outputs were used to include, exclude, rank, prioritize, flag records, or determine what moved to full-text retrieval or later review stages. Report the number of records reviewed by humans, LLMs, or both; the audit procedure for false exclusions; disagreement handling; thresholds if used; and final human responsibility. \\
LLM-M8 - Data extraction workflow & For LLM-assisted extraction, report the extraction schema, fields extracted, evidence sources used, output format, validation sample, manual verification rate, adjudication process, schema revisions or edge-case rules, and how missing, ambiguous, conflicting, tabular, or supplementary evidence was handled. \\
LLM-M9 - Annotation, harmonization, and LLM-as-judge use & For annotation, harmonization, quality assessment, metric extraction, or LLM-as-judge workflows, report label definitions, rubrics, judge model or version where applicable, calibration examples or samples, human calibration labels, reference standards, human audit procedures, agreement checks, and whether LLM outputs were final evidence or intermediate assistance. \\
LLM-M10 - Human oversight and adjudication & Describe human interaction with LLM outputs at each stage: who reviewed outputs, what proportion was checked, whether review was independent, how corrections were made, how disagreements were resolved, and who had final responsibility for review decisions. \\
LLM-M11 - Evaluation plan & State the planned evaluation dimensions using the \ouralgo reporting dimensions: performance, comparisons, modifications or optimization, resources and feasibility, and limitations or failure modes. Specify the minimum coverage implied by the implementation level and any stronger evaluation required by task consequence, limited human verification, system opacity, or irreversibility. \\
LLM-M12 - Data governance and reproducibility plan & Describe how input, output, intermediate data, prompts, schemas, code, audit logs, and validation materials were stored and managed. Report privacy, copyright, terms-of-service, vendor-access, or data-sharing constraints that limit reproducibility. \\
\addlinespace[0.35em]

\multicolumn{2}{@{}l}{\textit{Results}} \\
LLM-R1 - AI and human processing counts & Report how many records, reports, fields, annotations, claims, judgments, or outputs were processed by LLMs or software, by humans, or by both. Where screening was automated or prioritized, distinguish human and LLM/software decisions in the text or flow diagram. \\
LLM-R2 - Evaluation results & Report evaluation results across the dimensions required for the workflow complexity level. Include task-specific performance evidence, comparative evidence, optimization or modification results, resource or feasibility evidence, and limitation or failure-mode evidence as applicable. \\
LLM-R3 - Errors, disagreements, and corrections & Report observed false positives, false negatives, hallucinations, failed extractions, parsing failures, prompt failures, disagreement patterns, corrected outputs, and audit outcomes. Describe how these errors affected the review workflow or final dataset. \\
LLM-R4 - Resource and feasibility outcomes & If the review claims efficiency, scalability, workload reduction, lower cost, or improved feasibility, report the evidence supporting those claims, such as human time, compute cost, API cost, runtime, number of records handled, or implementation burden. \\
\addlinespace[0.35em]

\multicolumn{2}{@{}l}{\textit{Discussion}} \\
LLM-D1 - Limitations and failure modes & Discuss limitations introduced by LLM or software use, including prompt sensitivity, model or version dependence, hallucination, extraction errors, missed records, validation limits, proprietary opacity, data-quality problems, and remaining human workload. Explain how these limitations may have affected the review process, evidence base, extracted data, synthesis, or conclusions. \\
LLM-D2 - Experience and implications & Discuss what was learned from using the LLM workflow, including what worked, what failed, what required human judgment, and what future reviewers should know before reusing a similar approach. \\
\addlinespace[0.35em]

\multicolumn{2}{@{}l}{\textit{Other information}} \\
LLM-O1 - Availability of materials & Provide stable record identifiers, prompts, system instructions, model and tool versions, settings, schemas, code, screening or extraction labels when applicable, validation samples, audit logs, raw and parsed outputs, annotation guidelines, calibration artifacts, and reproducibility materials when legally and ethically possible. \\
LLM-O2 - Support, access, and competing interests & Report material support, sponsored access, credits, private model access, vendor involvement, or competing interests related to the LLM or software systems used in the review. \\
\bottomrule
\noalign{\vskip 7pt}
\caption{\textbf{Expanded \ouralgo checklist for LLM-assisted and software-mediated systematic reviews.} Items are organized by the corresponding PRISMA 2020 manuscript sections and expand the page-sized main-text checklist with task-specific detail. The framework is intended for use alongside, not instead of, PRISMA 2020.}
\label{tab:prisma-llm-checklist}\\
\end{longtable}
\endgroup

\begin{table}[H]
\centering
\footnotesize
\setlength{\tabcolsep}{4pt}
\renewcommand{\arraystretch}{1.15}
\begin{tabular}{@{}C{0.08\textwidth}L{0.22\textwidth}L{0.27\textwidth}L{0.35\textwidth}@{}}
\toprule
Level & Workflow type & Typical examples & Minimum evaluation reporting \\
\midrule
1 & Off-the-shelf tool or software & Named review software, AI-enabled search tools, or product interfaces used without custom prompting or engineering. & Limitations or failure modes and validation constraints. Add task-specific performance evidence whenever outputs can alter the evidence base. \\
2 & Prompt-only LLM & Direct prompting for screening, extraction, summarization, annotation, or drafting without few-shot examples or additional workflow engineering. & Level 1 plus task-specific performance evidence for the automated task. \\
3 & Few-shot, structured, or engineered workflow & Few-shot prompting, structured outputs, JSON extraction, rubrics, batching, prompt chains, voting, ensembles, or LLM-as-judge procedures. & Level 2 plus comparative evidence and evidence about consequential prompt, schema, rubric, ensemble, or workflow modifications. \\
4 & Retrieval-augmented, adapted, or single-agent workflow & RAG over records or PDFs, embeddings, retrieval pipelines, fine-tuned or otherwise adapted models, or a single agent operating over review materials. & All five dimensions: performance, comparisons, modifications or optimization, resources and feasibility, and limitations or failure modes. \\
5 & Multi-agent or swarm workflow & Multiple coordinated agents that divide, check, debate, retrieve, extract, judge, or synthesize review evidence. & All five dimensions plus coordination-specific failures, disagreement resolution, and evidence for any claimed benefit of cross-agent checking. \\
\bottomrule
\end{tabular}
\par\vspace{7pt}
\caption{\textbf{\ouralgo implementation levels and minimum evaluation-reporting expectations.} The levels organize increasing implementation complexity and are not intended as a risk score. Minimum expectations accumulate across levels, but consequential automation, limited human verification, proprietary or mutable systems, or irreversible decisions can justify stronger evaluation at any level. Reporting richness is retained as a descriptive empirical index in this study rather than a compliance threshold.}
\label{tab:prisma-llm-complexity}
\end{table}

\end{document}